\documentclass[aps,prl,twocolumn]{revtex4-1}

\newcommand{\br}{\boldsymbol{\textbf{r}}}

\newcommand{\dr}{\,d\br}

\newcommand{\vxc}{v_{\text{xc}}}
\newcommand{\rhod}{\rho_{\text{data}}}
\newcommand{\vFA}{v_{\text{FA}}}

\usepackage{multibib}
\newcites{supp}{Supplementary References}
\usepackage{mathtools}
\usepackage{hyperref}
\usepackage{subfiles}
\usepackage{graphicx}
\usepackage{graphics}
\usepackage{multirow}
\usepackage{fancyhdr, ifpdf}
\usepackage{amsxtra}
\usepackage{stmaryrd}
\usepackage{mathrsfs}
\usepackage{psfrag}
\usepackage{subfigure}
\usepackage{epsfig}
\usepackage{rotating}
\usepackage{setspace}
\usepackage{float}
\usepackage{amsfonts}
\usepackage{amsbsy}
\usepackage{amscd}
\usepackage{amsthm}
\usepackage{color}
\usepackage{tabularx}
\usepackage{threeparttable}
\usepackage{booktabs}
\usepackage{sidecap}
\usepackage{dcolumn}
\usepackage{footnote}
\usepackage{algorithm}
\usepackage{algorithmic}
\usepackage[normalem]{ulem}
\usepackage[titletoc,title]{appendix}

\definecolor{hellgruen}{rgb}{0.2,0.7,0.2}

\newcolumntype{M}[1]{>{\centering\arraybackslash}m{#1}}
\newcolumntype{N}{@{}m{0pt}@{}}

\newcommand{\norm}[1]{\left\lVert#1\right\rVert}
\newcommand{\modulus}[1]{\left\vert#1\right\vert}

\begin{document}
\title{A comparison of exact and model exchange-correlation potentials for molecules}
\author{Bikash Kanungo}
\affiliation{Department of Mechanical Engineering, University of Michigan, Ann Arbor, Michigan 48109, USA}
\author{Paul M. Zimmerman}
\affiliation{Department of Chemistry, University of Michigan, Ann Arbor, Michigan 48109, USA}
\author{Vikram Gavini}
\affiliation{Department of Mechanical Engineering, University of Michigan, Ann Arbor, Michigan 48109, USA}
\affiliation{Department of Materials Science and Engineering, University of Michigan, Ann Arbor, Michigan 48109, USA}

\begin{abstract}
Accurate exchange-correlation (XC) potentials for 3-dimensional systems---via solution of the \emph{inverse} density functional theory (DFT) problem---are now available to test the quality of DFT approximations. 
Herein, the \emph{exact} XC potential for six molecules---hydrogen molecule at three different bond-lengths, lithium hydride, water, and ortho-benzyne---are computed using accurate ground-state densities from full configuration interaction (CI) calculations. These potentials are then compared to model XC potentials obtained from DFT calculations with commonly used non-local (B3LYP, HSE06, SCAN0, and M08-HX) and local/semi-local (SCAN, PBE, PW92) XC functionals. While relative errors in the ground-state densities from these models are order $\mathcal{O}(10^{-3}-10^{-2})$, much larger errors in the model XC potentials are found, $\mathcal{O}(10^{-1}-10^0)$, in both the $L_2$ norm of the potential as well as its gradients. These errors are exacerbated in strongly correlated situations, such as the stretched $\text{H}_2$ molecule. Among the model XC functionals under consideration, SCAN0 offers the best quantitative and qualitative agreement with the exact XC potential, underlining the significance of satisfying the exact conditions as well as the incorporation of non-local effects in the construction of XC functionals. Overall, this work indicates that tests against the exact XC potential will provide a promising new direction for building more accurate XC functionals for DFT.

\end{abstract}

\maketitle

Density functional theory (DFT) has remained the most popular electronic structure theory for the past 40 years~\cite{Becke2014,jJones2015,Mardirossian2017}, owing to its great balance of speed and accuracy. Within the Kohn-Sham formalism~\cite{Hohenberg1964,Kohn1965}, DFT presents a formally exact reduction of the interacting many-electron Schr\"odinger equation to an equivalent problem of noninteracting electrons in an effective mean-field that is governed by the ground-state electron density ($\rho(\br)$). However, in practice, DFT has remained far from exact due to the unavailability of the exact exchange-correlation (XC) functional that encapsulates the quantum many-electron interactions into a mean-field dependent on $\rho(\br)$. Traditionally, DFT calculations have employed local XC functionals defined in terms of the density (e.g., local density approximations (LDA)~\cite{Perdew1992}) and its gradients (e.g., generalized gradient approximation (GGA)~\cite{Perdew1996}). In the past three decades, DFT has witnessed a growing popularity of two increasingly sophisticated families of XC functionals: (a) the meta-GGA, which includes an additional dependence on the kinetic energy density (e.g., SCAN~\cite{Sun2015}, TPSS~\cite{Tao2003}), and (b) the hybrid XC functionals, wherein LDA or GGA exchange is mixed with a fraction of Hartree-Fock exchange (e.g., B3LYP~\cite{Becke1993a,Becke1993b,Lee1988}, HSE06~\cite{Heyd2003,Paier2006}, PBE0~\cite{Adamo1999}). Unlike the LDA or the GGA, the meta-GGA and the hybrid functionals entail the use of the Kohn-Sham orbitals, thereby leading to a nonmultiplicative potential in the Kohn-Sham equations. Formally, the theoretical basis for the use of the meta-GGA and hybrid functionals have been established through the generalized Kohn-Sham (GKS) formalism~\cite{Seidl1996}. In practical terms, meta-GGAs and hybrids have greatly enhanced the predictive capability of DFT for molecules~\cite{Cohen2000,Kummel2008, Sun2016} and solids~\cite{Distasio2014,Marsman2008, Isaacs2018}.

Despite the incredible success of the hybrid XC functionals, fundamental deficiencies still persist---self-interaction, delocalization, and static correlation errors, to name a few~\cite{Cohen2012}. Thus, development of XC functionals that provide satisfactory accuracy for weakly and strongly correlated systems still remains a serious challenge in DFT. To that end, the inverse DFT approach~\cite{Zhao1994, Leeuwen1994, Peirs2003, Wu2003, Jensen2018, Kanungo2019, Shi2021} of finding the XC potential ($\vxc(\br)$) corresponding to a given density ($\rhod(\br)$) provides an instructive tool in investigating the deficiencies of existing model XC functionals. To elaborate, a comparison of the XC potentials corresponding to ground-state densities from meta-GGA or hybrid XC based GKS calculation against the \emph{exact} XC potential for the ground-state density obtained from accurate many-body calculation (e.g., configuration interaction (CI)) can inform the features missed by the approximate XC functionals, and in turn, assist in designing next generation functionals. For ease of reference, we term the XC potentials corresponding to the densities from approximate XC functionals as \emph{model} XC potentials. In the past, a few such comparative studies have been conducted via the optimized effective potential (OEP) approach~\cite{Kummel2008}, wherein one finds a local potential that minimizes an orbital-dependent energy functional~\cite{Grabowski2011,Smiga2020}.  
However, unlike the unified approach in inverse DFT of evaluating the model and the exact XC potentials from their respective densities, the OEP-based approach treats the OEP and the exact XC potential differently (i.e., the OEP is evaluated by minimizing an energy functional and the exact XC potential is obtained to yield the CI density).  

Given the importance of inverse DFT in assessing and developing XC functionals, over the last three decades, several approaches have been developed to solve the inverse DFT problem~\cite{Shi2021}. While analytic evaluation of the $\vxc$ from the density is feasible for one- or two-electron systems~\cite{Umrigar1994, Ryabinkin2017}, systems with more than two-electrons warrant numerical evaluation.    
Broadly speaking, the numerical approaches to inverse DFT can be categorized as an iterative procedure~\cite{Gorling1992, Wang1993, Leeuwen1994, Peirs2003, Ryabinkin2012} or a constrained optimization~\cite{Zhao1994, Tozer1996, Wu2003, Jacob2011, Kanungo2019, Garrick2020}. Most of the approaches have suffered numerical artifacts which manifest as spurious oscillations in the resultant $\vxc$ and/or as non-unique solutions. One primary source of these artifacts is the incompleteness of the Gaussian basis used to discretize the problem~\cite{Burgess2007, Bulat2007,Jacob2011}. In other words, the problem is well-posed only in a complete basis. Various attempts have been made to mitigate spurious oscillations, for instance through a combination of regularization and truncated singular value decomposition (TSVD)~\cite{Burgess2007, Jacob2011}. When target densities are computed using a Gaussian basis, artifacts also arise due to incorrect asymptotics. Specifically, the Gaussian basis-set based densities lack the cusp at the nuclei and lack the proper exponential decay at long distances. These incorrect asymptotics induce large unphysical oscillations in the resultant $\vxc$~\cite{Mura1997, Schipper1997, Gaiduk2013, Kanungo2019}. For single atoms and ions, wherein the 3D equations can be reduced to 1D owing to spherical symmetry, several efforts ~\cite{Leeuwen1994,Schipper1997,Peirs2003,Ryabinkin2012,Garrick2020} have circumvented the artifacts related to Gaussian basis-set based densities by using a Slater basis or a radial grid. Extension of these techniques to molecules (3D systems) remains computationally challenging, due to the difficulty in employing a Slater basis or a grid for CI calculations involving molecules. A substantially different approach~\cite{Ryabinkin2015, Cuevas2015, Ospadov2017}, utilizing the two-electron reduced density matrix (2-RDM), has also been proposed to remedy the non-uniqueness and the spurious oscillations in the obtained $\vxc$, and is considered one of the state-of-the-art methods in obtaining the exact $\vxc$. When using a finite basis, however, this approach need not correspond to the solution of the inverse DFT problem, as the provided $\vxc$ is not guaranteed to yield the target electron density~\cite{Ospadov2017}. An extension of this approach using multi-resolution analysis (MRA), a complete basis, has been proposed to construct the exchange potential from Hartree-Fock densities~\cite{Stuckrath2021}. The capabilities and accuracy afforded by some of these important approaches to solve the inverse DFT problem are given in Table~\ref{tab:comparison}.


\begin{table*} 
  \caption{Comparison of inverse DFT approaches from literature, in terms of level of theory used to obtain the target density $\rhod$, the basis used for Kohn-Sham orbitals, the largest system considered, and the range of errors in the density. The comparison only includes methods for which errors in density for molecular (3D) systems are reported. $err_{\rho}=\frac{1}{N_e}\int|\rhod-\rho|\dr$ is the $L_1$ error in density normalized with the number of electrons.} 
  \label{tab:comparison}
  \begin{threeparttable} 
  \begin{tabular}{| M{3cm} | M{2cm} | M{4cm} | M{2.55cm} | M{4cm} |}
    \hline
    Method & Theory & Basis & Largest System (electrons) & $err_\rho$ \\
    \hline
    \hline
    KZG~\cite{Kanungo2019} & iFCI & finite element\tnote{*} & $\text{C}_6\text{H}_4$ (40) & $3.4\times10^{-5}-8.2\times10^{-5}$ \\ \hline
    modified-RKS~\cite{Ospadov2017} & CASSCF & gaussian/cc-pCV5Z\tnote{**} & HCN (14) & $5\times10^{-4}$ \\ \hline
    SGB~\cite{Schipper1998} & MRCI & gaussian/cc-pCVTZ\tnote{**} & $\text{F}_2$ (18) &  $3.8\times10^{-4}-6.7\times10^{-3}$ \\ \hline
    DCEP-MRA~\cite{Stuckrath2021} & HF & MRA\tnote{*} &  $\text{C}_5\text{H}_5\text{N}_5$ (70) & $2.8\times10^{-6}-4.2\times10^{-6}$\tnote{$\dagger$} \\ \hline
  \end{tabular}
   \begin{tablenotes}
   \item[*] complete basis 
   \item[**] incomplete basis 
   \item [$\dagger$] the calculation for the target density ($\rhod$) and the inverse DFT problem are done in the same basis, which allows for better accuracies in the density (see Supplementary Table 1 in~\cite{Kanungo2019})
   \end{tablenotes}
  \end{threeparttable}
\end{table*}

Given the various numerical shortcomings of inverse DFT approaches for molecules (3D systems), a systematic comparative study of the exact and model XC potentials for polyatomic systems, solved to chemical accuracy, is lacking. The present study aims to address this important gap.
The main aspect enabling this study is the recent development of numerical approaches that provide a robust and accurate solution to the inverse DFT problem that resolves the aforementioned outstanding challenges in inverse DFT. 
In a recent effort~\cite{Kanungo2019}, building on prior works~\cite{Wu2003,Gaiduk2013, Jensen2018}, we proposed an approach to overcome these numerical issues, and demonstrated an accurate solution to the inverse DFT problem for polyatomic systems that include weakly and strongly correlated systems. In this study, we adopt this recent development to conduct a comparative study of the exact and model XC potentials. 

Given an electron density $\rhod(\br)$, the inverse DFT problem of finding the $\vxc(\br)$ that yields the density can be posed as partial differential equation (PDE) constrained optimization:  
\begin{equation} \label{eq:Inverse}
\text{arg}\min_{\vxc(\br)}\int{w(\br)\left(\rhod(\br)-\rho(\br)\right)^2\dr}\,,
\end{equation}
subject to:
\begin{equation}\label{eq:Forward} 
    \left(-\frac{1}{2}\nabla^2+v_{\text{ext}}(\br)+v_{\text{H}}(\br)+\vxc(\br)\right)\psi_i=\epsilon_i\psi_i\,,
\end{equation}
\begin{equation} \label{eq:Normal}
\int{|\psi_i(\br)|^2\dr} =1\,. 
\end{equation}
In the above equation, $w(\br)$ is an appropriately chosen positive weight to expedite convergence; $v_{\text{ext}}(\br)$ represents the nuclear potential; $v_{\text{H}}(\br)$ is the Hartree potential corresponding to $\rhod(\br)$; and $\psi_i$ and $\epsilon_i$ denote the Kohn-Sham orbitals and eigenvalues, respectively. For simplicity, we restrict ourselves to closed-shell systems, and, hence, the Kohn-Sham density $\rho(\br) = 2\sum_{i=1}^{N_e/2}{|\psi_i(\br)|^2}$. We employ an adjoint state approach to solve the above PDE-constrained optimization, and refer to~\cite{Kanungo2019} for details of the formulation. We discretize the $\psi_i$'s using an adaptively refined, fourth-order spectral finite-element (FE) basis. The $\vxc$, on the other hand, is discretized using linear FE basis owing to the smoother nature of the XC potential. The completeness of the FE basis is crucial to obtaining an accurate solution to the inverse DFT problem. 

In order to mitigate the unphysical behavior in $\vxc$ induced from incorrect asymptotics in densities computed using Gaussian basis-sets, as is typically in CI calculations or DFT calculations with hybrid functionals, we employ the following two strategies. First, to fix the incorrect asymptotics in the Gaussian density at the nuclei, we add a small correction to $\rhod(\br)$ given by $\Delta\rho(\br)=\rho_{\text{FE}}^{\text{DFT}}(\br)-\rho_{\text{G}}^{\text{DFT}}(\br)\,$. Here $\rho_{\text{FE}}^{\text{DFT}}(\br)$ is the ground-state density obtained using a known XC functional (e.g., LDA, GGA) in the FE basis and $\rho_{\text{G}}^{\text{DFT}}(\br)$ denotes the same, except using a Gaussian basis. The $\Delta\rho$ correction accounts for the basis set error in the Gaussian density around the nuclei (cf.~\cite{Kanungo2019}).  
Secondly, appropriate boundary conditions on $\vxc$ are enforced in the low density region ($\rhod < 10^{-7}$) to alleviate the numerical artifacts stemming from the incorrect far-field decay of the Gaussian densities.
In particular, we use a scaled Fermi-Amaldi potential~\cite{Ayers2005}, given by $\vFA(\br)=-\frac{\alpha}{N_e}\int\frac{\rhod(\br')}{|\br-\br'|}\dr'$, as the boundary condition. In the evaluation of \emph{exact} $\vxc$ potentials (corresponding to CI densities), we choose $\alpha=1$. Whereas, for the $\vxc$ potentials corresponding to densities from hybrid-DFT calculations, $\alpha$ is set to the fraction of the Hartree-Fock exchange used in hybrid XC functional, ensuring the consistent far-field decay of the $\vxc$. In the case of densities obtained using SCAN functionals, we use the Slater exchange potential, $v_{\text{S}}(\br)=-{\left(\frac{3}{\pi}\right)}^{1/3}\rhod^{1/3}(\br)$, as the boundary condition.

To compare exact and model $\vxc$ potentials, six molecules were chosen as benchmark systems. These include the hydrogen ($\text{H}_2$) molecule with three different bond lengths ($R_{\text{H-H}}$): (a) the equilibrium bond-length ($R_{\text{H-H}}$ = 1.4 a.u.\,), denoted as $\text{H}_2(eq)$; (b) $\sim0.8\times$ the equilibrium bond-length  ($R_{\text{H-H}}$ = 1.13 a.u.\,), denoted as $\text{H}_2(0.8eq)$; and (c) $\sim2\times$ the equilibrium bond-length ($R_{\text{H-H}}$ = 2.84 a.u.\,), denoted as $\text{H}_2(2eq)$. Additionally, we also use lithium hydride (LiH) molecule ($R_{\text{Li-H}}$ = 3.01 a.u.), water ($\text{H}_2\text{O}$) molecule ($R_{\text{O-H}}$= 1.89 a.u. and $\text{H-O-H}$ angle of $104.11^{\circ}$), and ortho-benzyne ($\text{C}_6\text{H}_4$) molecule as benchmark systems. 

The exact $\vxc$ potentials for these benchmark systems are obtained from inverse DFT calculations using accurate ground-state densities from incremental full CI (iFCI) calculations~\cite{Zimmerman2017,Zimmerman2017b}. The model $\vxc$ potentials are similarly obtained from DFT ground-state densities of several widely used approximate XC functionals: (i) B3LYP~\cite{Becke1993b,Lee1988}, a widely used hybrid GGA; (ii) HSE06~\cite{Heyd2003,Paier2006}, a popular range-separated hybrid GGA; (iii) SCAN~\cite{Sun2015}, a meta-GGA designed to satisfy several of the known exact conditions on the XC functional; (iv) SCAN0~\cite{Hui2016}, a recently developed hybrid extension of SCAN; (v) M08-HX~\cite{Zhao2008}, a semi-empirical hybrid meta-GGA functional. Additionally, we present comparisons against $\vxc$ potentials corresponding to two historically significant XC functionals---PBE (a GGA)~\cite{Perdew1996} and PW92 (an LDA)~\cite{Perdew1992}.
All the iFCI and DFT calculations to compute ground-state densities are done using the QChem software package~\cite{QChem4} with a polarized, triple zeta Gaussian basis set (cc-pVTZ ~\cite{Dunning1989}). 
In all the inverse DFT calculations conducted, the $L_2$ norm error in the density---$\norm{\rhod-\rho}_{L_2}=\left(\int\left(\rhod(\br)-\rho(\br)\right)^2\dr)\right)^{1/2}$---is driven below $10^{-4}$. 

In order to quantify the difference between the exact and the model $\vxc$ potentials, we use four different error metrics
\begin{subequations} \label{eq:vxcErrors}
\begin{equation}
e_1 = \frac{\norm{\rho_{\text{data}}\delta\vxc}_{L_2}}{\norm{\rho_{\text{data}}\vxc^{\text{exact}}}_{L_2}}\,,
\quad e_2 = \frac{\norm{\rhod\modulus{\nabla\delta\vxc}}_{L_2}}{\norm{\rhod\modulus{\nabla\vxc^{\text{exact}}}}_{L_2}}\,,
\end{equation}
\begin{equation}
e_3 = \frac{\norm{\delta\vxc}_{L_2}}{\norm{\vxc^{\text{exact}}}_{L_2}}\,,
\quad e_4 = \frac{\norm{\modulus{\nabla\delta\vxc}}_{L_2}}{\norm{\modulus{\nabla\vxc^{\text{exact}}}}_{L_2}}\,,
\end{equation}
\end{subequations}
where $\delta\vxc=\vxc^{\text{exact}}-\vxc^{\text{model}}$. The error metric $e_2$ and $e_4$ are insensitive to any constant shift in the model $\vxc$ due to incorrect far-field asymptotics. We use the $\rhod$ weight in $e_1$ and $e_2$ to inform the errors in the model $\vxc$ in the regions that are energetically important. Table~\ref{tab:L2DiffWeight} presents the $e_1$ and $e_2$ errors for all the models across the five benchmark systems considered in this study. Unweighted error metrics ($e_3$ and $e_4$) are in the Supplementary Material.


Figure~\ref{fig:VXC_H2eq} compares the exact and the model $\vxc$ potentials for the $\text{H}_2(eq)$ molecule along its bond axis. As expected, all model $\vxc$ potentials differ significantly from the exact one in the low density region on account of incorrect far-field asymptotics. Importantly, the results also indicate large errors for model $\vxc$ potentials in the bonding region. 
The $\vxc$ corresponding to M08-HX exhibits large oscillations, likely due to serious errors in treatment of the exchange potential~\cite{Mardirossian2013,Medvedev2017}. Qualitatively, $\vxc$ from SCAN and SCAN0 provide the closest resemblance with the exact $\vxc$. 
Error metrics ($e_1$-$e_4$) also suggest that SCAN0 offers the best model for $\text{H}_2(eq)$ (see Table~\ref{tab:L2DiffWeight}). However, even for SCAN0, the errors in $\vxc$ are in the range of $\mathcal{O}(10^{-1}-10^{0})$, suggesting that there is a large scope for improvement in XC functionals. Furthermore, despite the density errors being $\mathcal{O}(10^{-3}-10^{-2})$---$\norm{\rhod^{\text{exact}}-\rhod^{\text{model}}}_{L_2}/\norm{\rhod^{\text{exact}}}_{L_2} \sim \mathcal{O}(10^{-3}-10^{-2})$ (cf. Supplementary Material for the density errors)---the corresponding error in $\vxc$ is $\mathcal{O}(10^{-1}-10^{0})$. Thus, it may be instructive to also use XC potentials---which exhibit greater sensitivity in comparison to the densities---in the design and development of future XC functionals.

\begin{figure}[htpb] 
    \centering
        \includegraphics[scale=0.7]{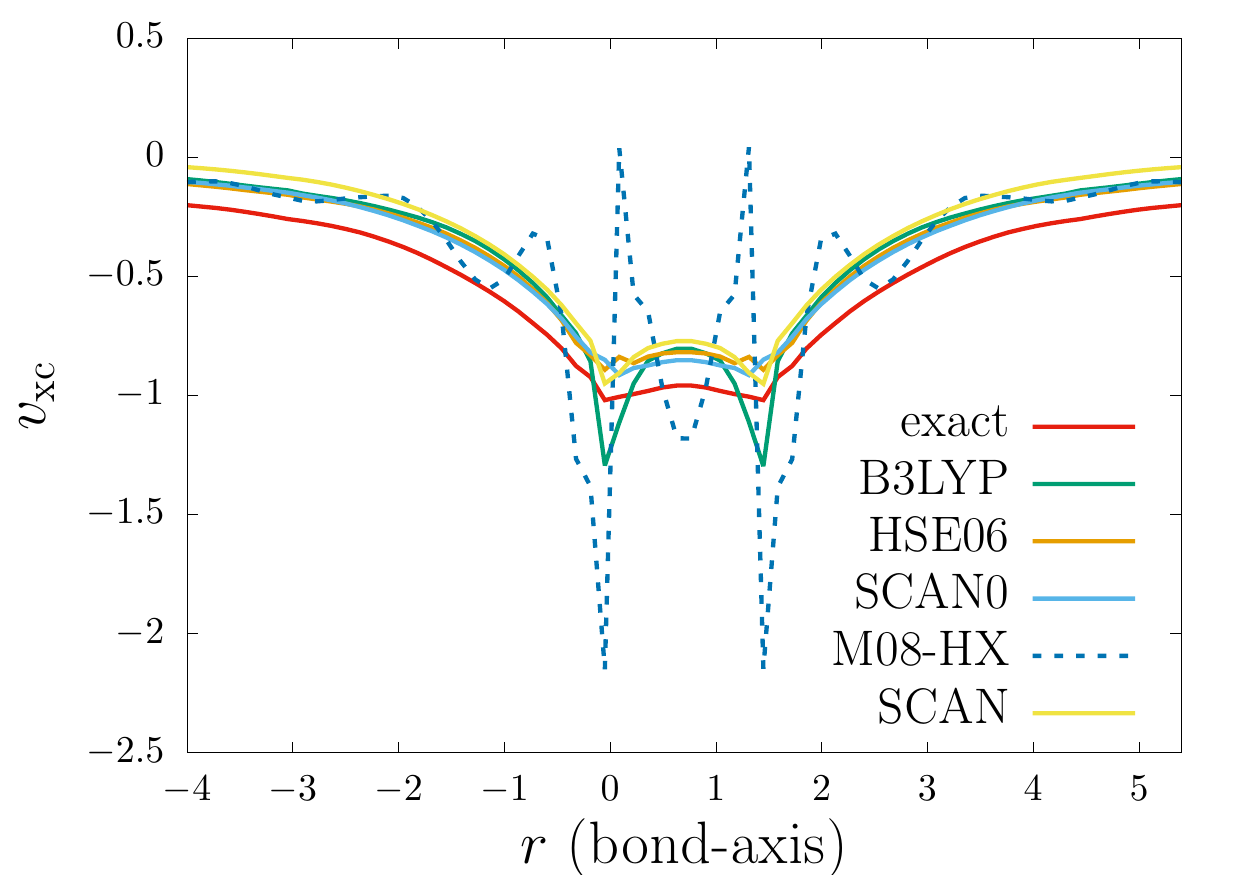}
		\caption{\small Comparison of the exact and model $\vxc$ for $\text{H}_2(eq)$. The H atoms are located at $r=0$ and $r=1.398$ a.u.\,. See Supplementary Material for comparison of the exact $\vxc$ with the PBE and PW92 based model $\vxc$ potentials.}
    \label{fig:VXC_H2eq}
\end{figure}



\begin{figure}[htpb] 
    \centering
        \includegraphics[scale=0.7]{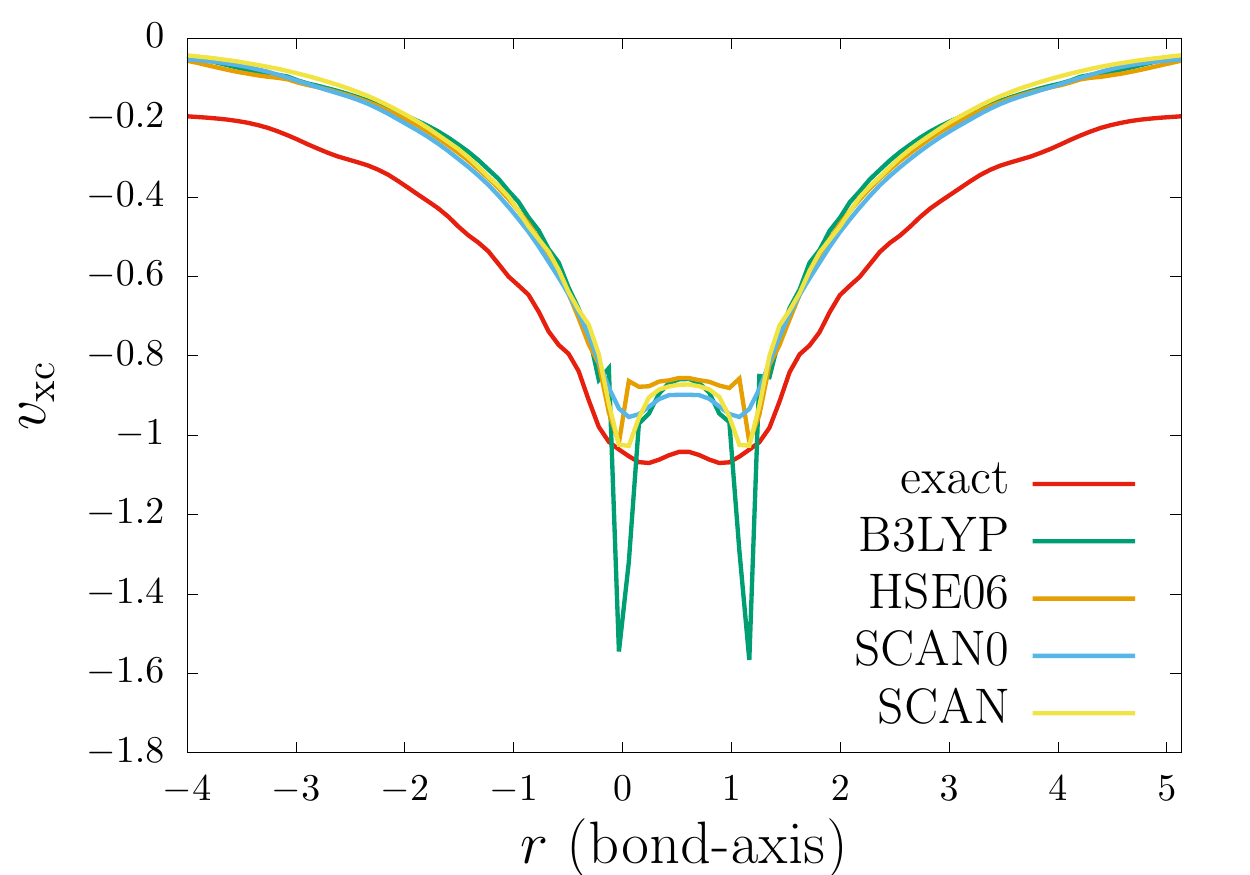}
		\caption{\small Comparison of the exact and model $\vxc$ for $\text{H}_2(0.8eq)$. The H atoms are located at $r=0$ and $r=1.134$ a.u.\,.}
    \label{fig:VXC_H20.8eq}
\end{figure}
\begin{figure}[htpb] 
    \centering
        \includegraphics[scale=0.7]{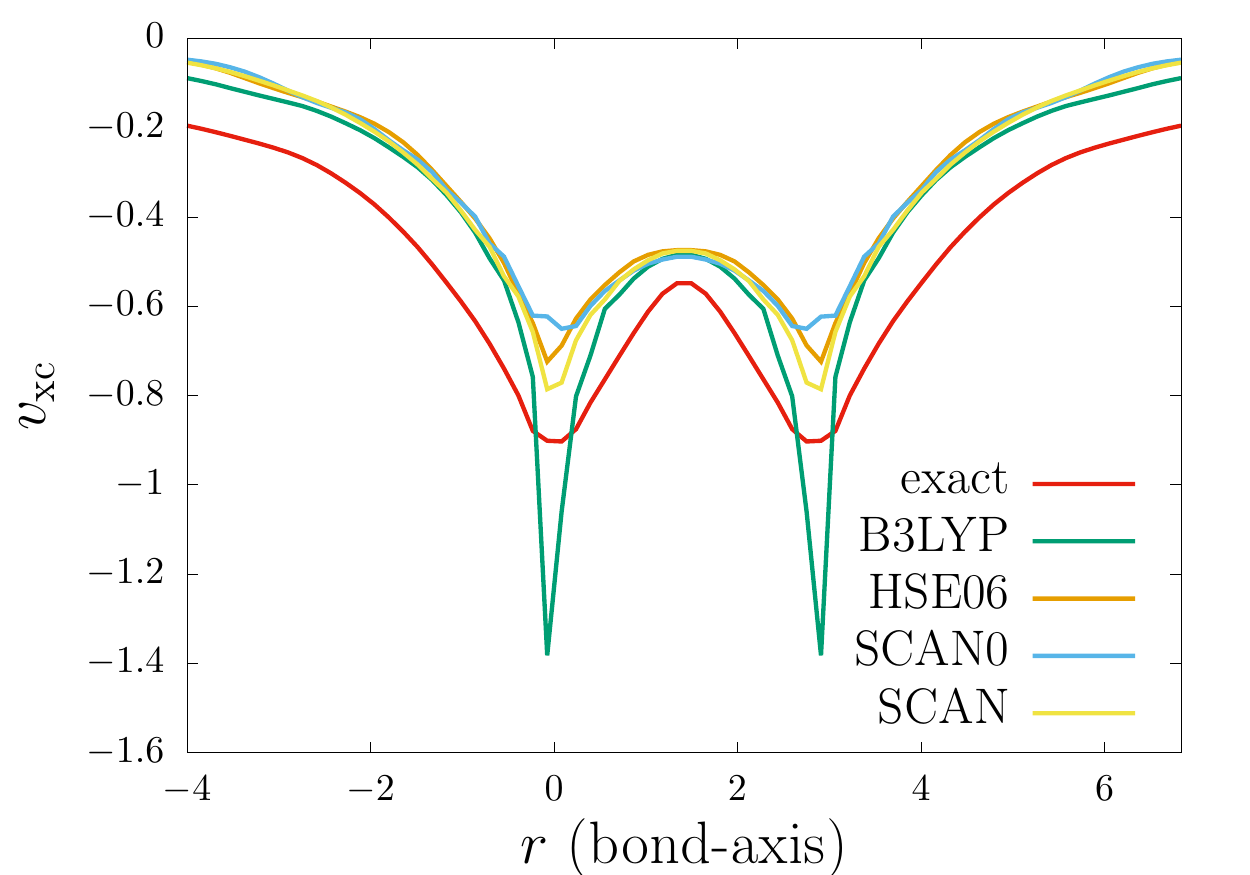}
		\caption{\small Comparison of the exact and model $\vxc$ for $\text{H}_2(2eq)$. The H atoms are located at $r=0$ and $r=2.835$ a.u.\,.}
    \label{fig:VXC_H22eq}
\end{figure}

For clarity in presenting the remaining molecules, a graphical comparison of the exact $\vxc$ potential against the B3LYP, HSE06, SCAN0, and SCAN model $\vxc$ potentials is provided (see the Supplementary Material for a comparison with the M08-HX, PBE, and PW92 model $\vxc$ potentials). Figures~\ref{fig:VXC_H20.8eq} \& ~\ref{fig:VXC_H22eq} present the comparison for $\text{H}_2(0.8eq)$ and $\text{H}_2(2eq)$ molecules, respectively. The $\text{H}_2(2eq)$ represents a prototypical case of strong electronic correlations, wherein all existing XC approximations perform poorly. As with the $\text{H}_2(eq)$ molecule, for $\text{H}_2(0.8eq)$ and $\text{H}_2(2eq)$, the model $\vxc$ potentials significantly differ from the exact $\vxc$ in the bonding region as well as in the far field. Further, in both cases, the $\vxc$ corresponding to B3LYP remains substantially deeper near the nuclei, owing to a greater concentration of electrons near the nuclei than the exact case. The SCAN0 based model $\vxc$ offers better qualitative and quantitative agreement with the exact $\vxc$ for $\text{H}_2(0.8eq)$, in comparison to other model $\vxc$ potentials. 
In the case of $\text{H}_2(2eq)$, the SCAN0, SCAN and HSE06 model $\vxc$ potentials exhibit similar structure and errors in relation to the exact $\vxc$. Quantitatively, SCAN0 exhibits larger errors for $\text{H}_2(2eq)$ than it does for $\text{H}_2$ and $\text{H}_2(0.8eq)$, suggesting a greater deficiency while handling strong correlations.


\begin{figure}[htpb] 
    \centering
        \includegraphics[scale=0.7]{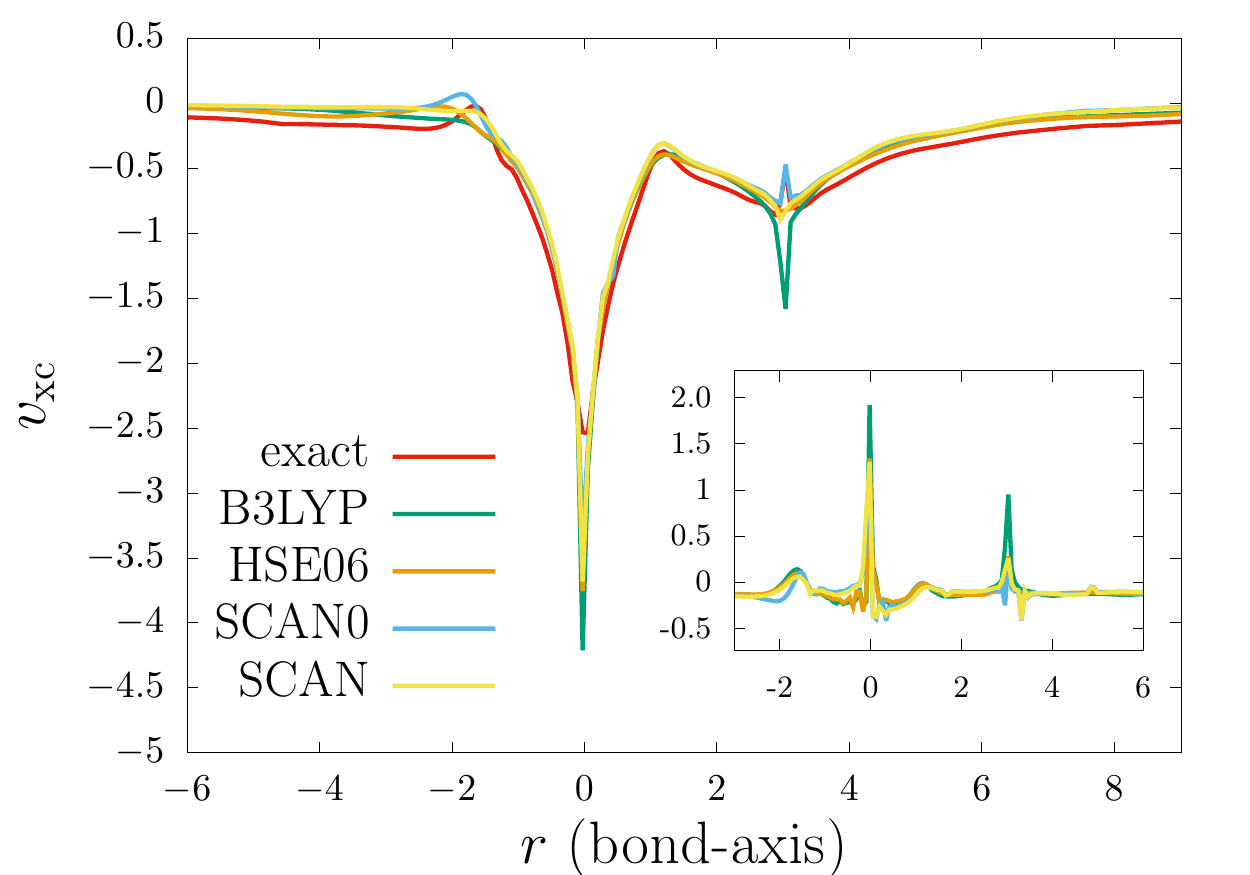}
		\caption{\small Comparison of the exact and model $\vxc$ for $\text{LiH}$. The inset shows $\delta\vxc=\vxc^{\text{exact}}-\vxc^{\text{model}}$. The Li and H atom are at $r=0$ and $r=3.014$ a.u.\,, respectively.}
    \label{fig:VXC_LiH}
\end{figure}

Figure~\ref{fig:VXC_LiH} compares the exact and model $\vxc$ potentials for the LiH molecule, along the bond axis. As evident, all the model $\vxc$ potentials are substantially deeper at the Li atom, suggesting a greater electronegativity on the Li atom than the exact one. There also is a distinct local maximum at the H atom in the exact $\vxc$ as well as SCAN0 based model $\vxc$, as opposed to a local minimum in all other $\vxc$ potentials. Importantly, both the exact and the SCAN0 based $\vxc$ exhibit an atomic inter-shell structure (near $r=1.2$ a.u.\, and $r=-1.6$ a.u.\,). The atomic inter-shell structure is a distinctive feature of the exact $\vxc$, which is typically absent in model $\vxc$ potentials, as is evident from the B3LYP, HSE06, and SCAN based $\vxc$. Comparing all model $\vxc$ potentials, SCAN0 based model $\vxc$ once again offers the best agreement with the exact $\vxc$, both qualitatively and quantitatively (cf. Table~\ref{tab:L2DiffWeight}).       

\begin{figure}[htpb] 
    \centering
        \includegraphics[scale=0.7]{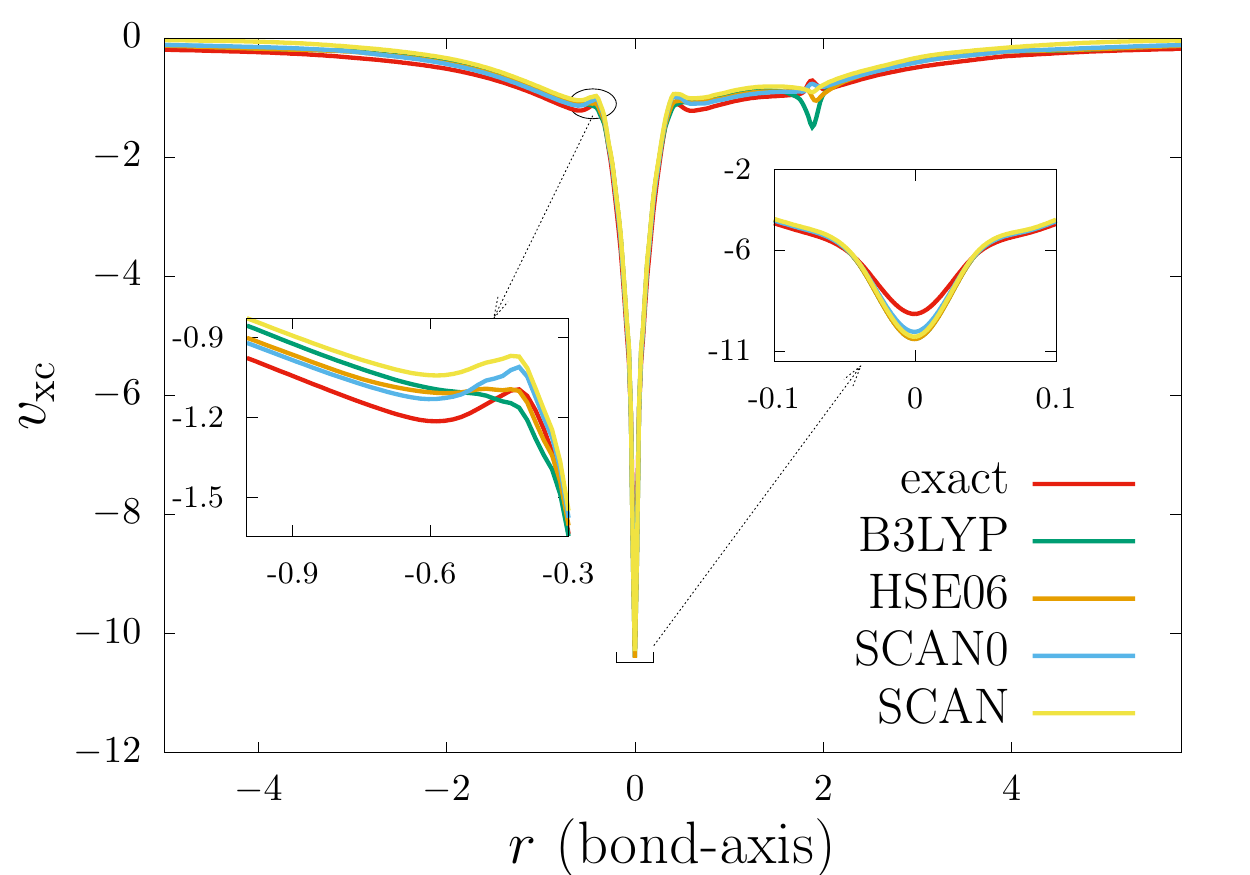}
		\caption{\small Comparison of the exact and model $\vxc$ for $\text{H}_2\text{O}$ along the O-H bond (cf. Supplementary Material for a similar comparison along the lone-pair axis). The O and H atom are at $r=0$ and $r=1.890$ a.u.\,, respectively. }
        \label{fig:VXC_H2O_bond}
\end{figure}
\begin{figure}[htbp]
\centering
\subfigure[]{ \label{fig:VXC_H2O_heatmap_exact_b3lyp}\includegraphics[scale=0.7]{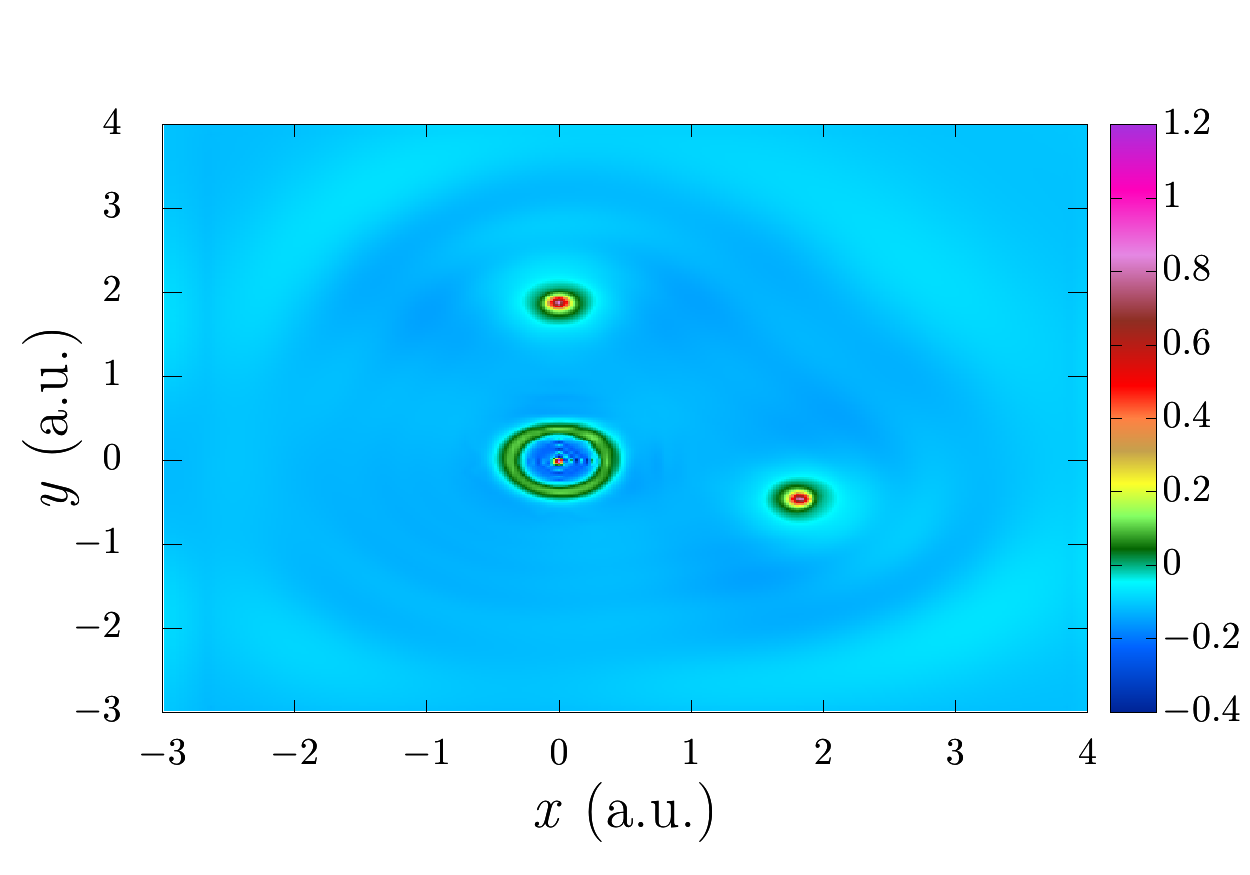}}
\vspace{-0.3cm}
\subfigure[]{ \label{fig:VXC_H2O_heatmap_exact_scan0}\includegraphics[scale=0.7]{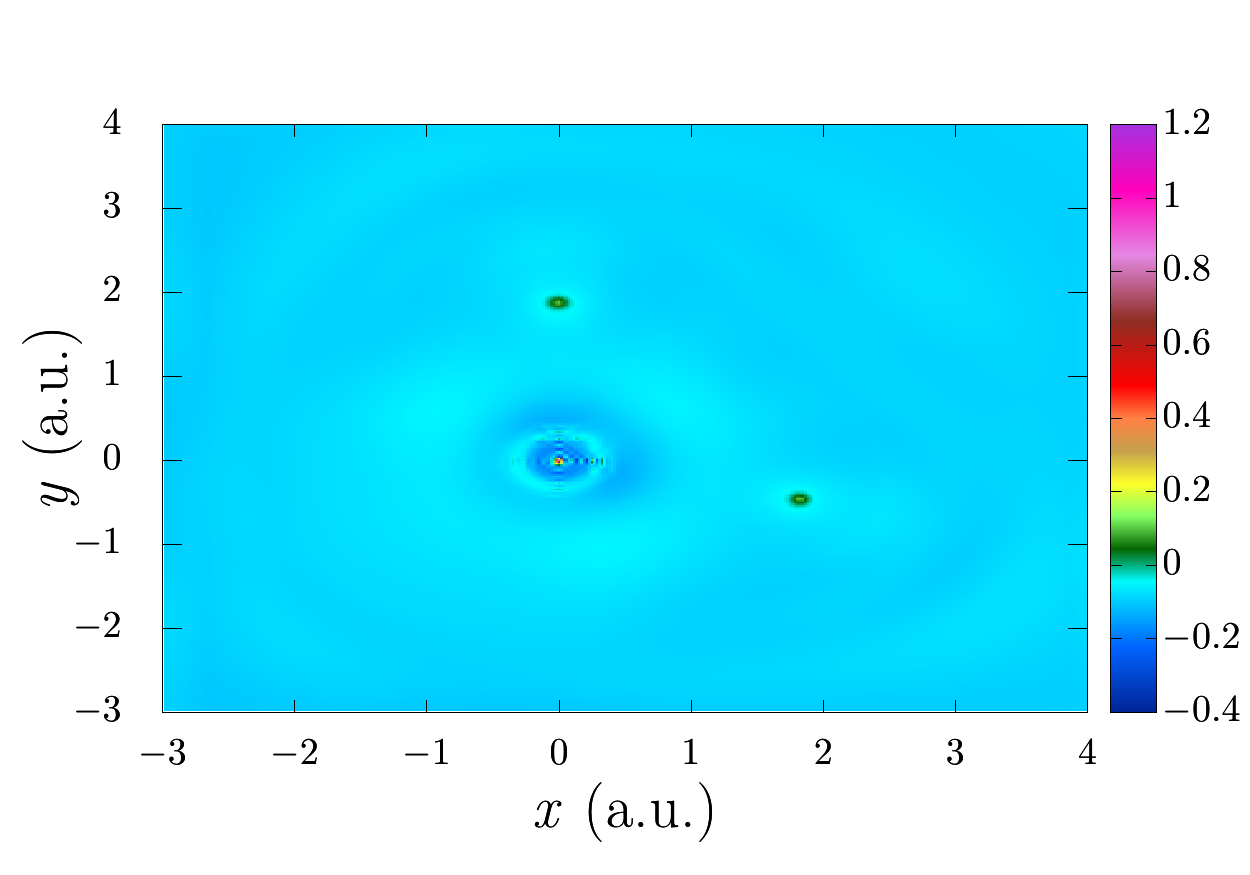}}
\caption{\small{Comparison of the exact and model $\vxc$ for $\text{H}_2\text{O}$ on the plane of the molecule: (a) difference between the exact and B3LYP based model $\vxc$, and (b) difference between the exact and SCAN0 based model $\vxc$.}}
\end{figure}

Figure~\ref{fig:VXC_H2O_bond} presents the exact and the model $\vxc$ potentials for $\text{H}_2\text{O}$ along an O-H bond. Errors in the molecular plane for B3LYP and SCAN0 model $\vxc$ (i.e., $\vxc^{\text{exact}}-\vxc^{\text{model}}$) are shown in Fig.~\ref{fig:VXC_H2O_heatmap_exact_b3lyp} \& Fig.~\ref{fig:VXC_H2O_heatmap_exact_scan0}, respectively. As is evident, all the model $\vxc$ potentials are too deep near $r=0$ (cf. inset in Fig.~\ref{fig:VXC_H2O_bond}), representing a higher electronegativity on the O atom compared to the exact $\vxc$. As with LiH, the exact $\vxc$ and the SCAN0 based model $\vxc$ feature a local maxima at the H atom, in contrast to a local minima for the other model $\vxc$ potentials. The exact $\vxc$ as well as the SCAN0 and SCAN $\vxc$ exhibit an atomic inter-shell structure around the O atom---marked by the local maxima and minima near $r=\pm0.4$ a.u.\, in Fig.~\ref{fig:VXC_H2O_bond}---otherwise absent in the B3LYP and HSE06 model $\vxc$.

\begin{figure}[htbp]
\centering
\subfigure[]{ \label{fig:VXC_benzyne_heatmap_exact_b3lyp}\includegraphics[scale=0.17]{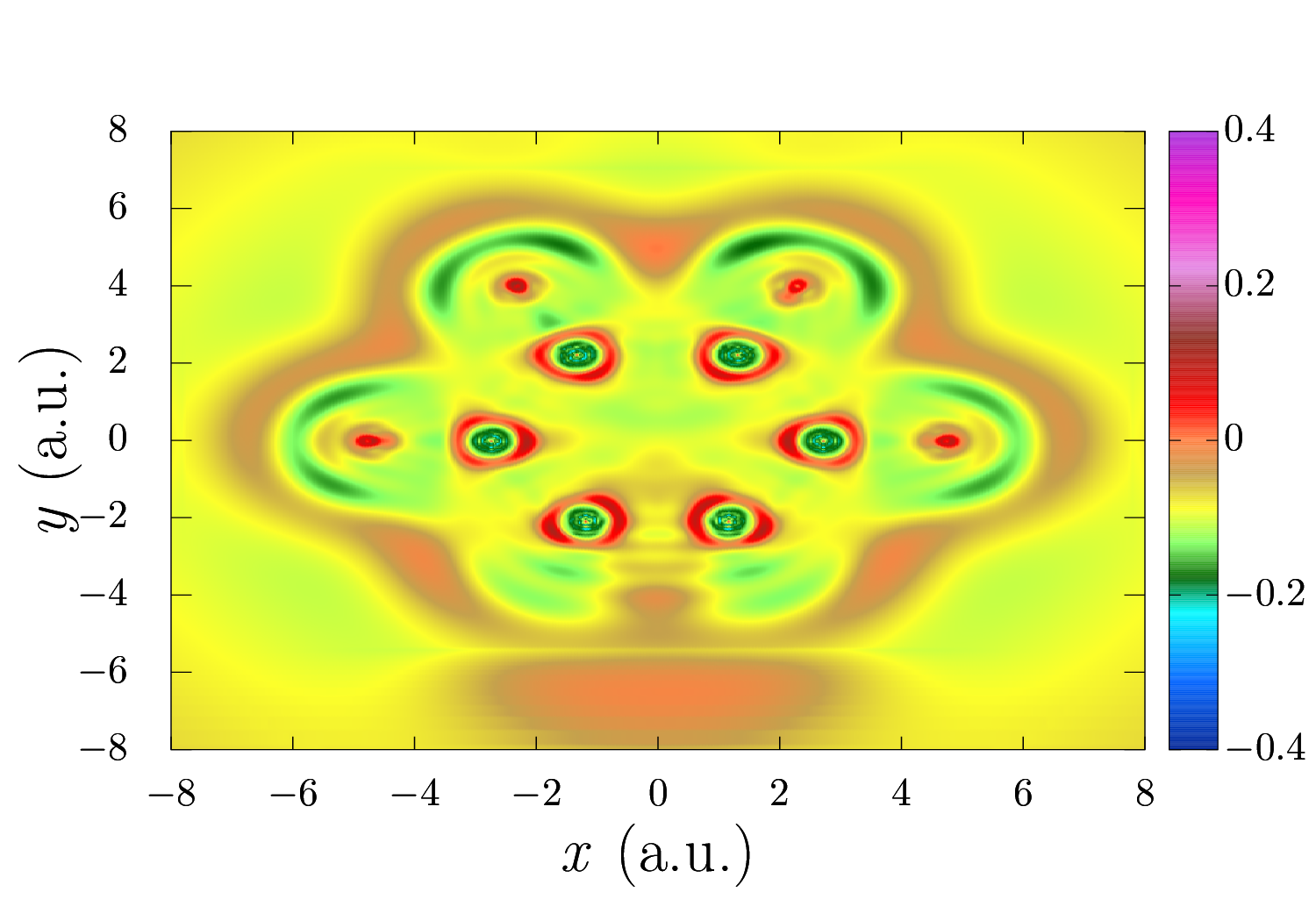}}
\vspace{-0.3cm}
\subfigure[]{ \label{fig:VXC_benzyne_heatmap_exact_scan0}\includegraphics[scale=0.17]{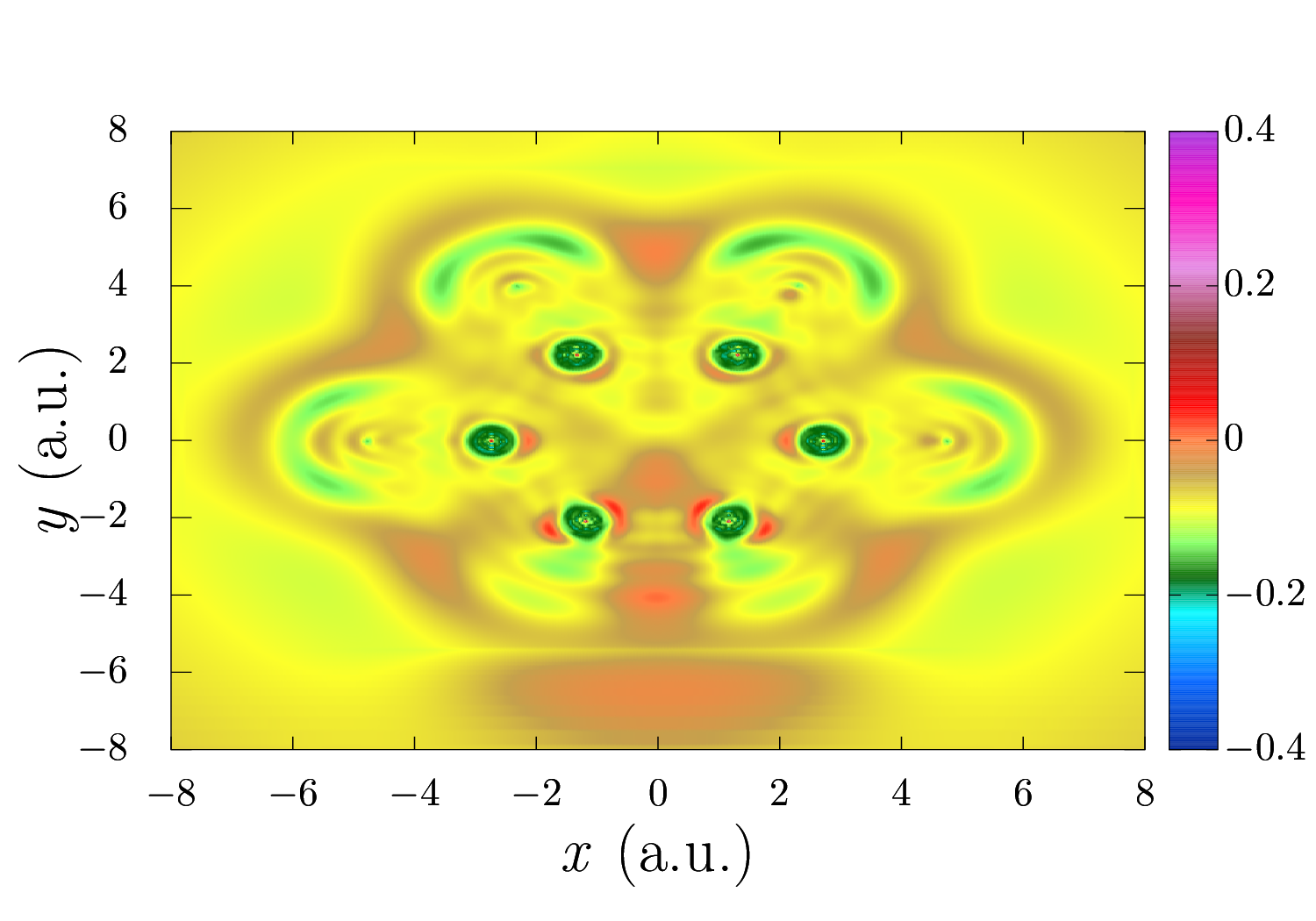}}
\caption{\small{Comparison of the exact and model $\vxc$ for $\text{C}_6\text{H}_4$ on the plane of the molecule: (a) difference between the exact and B3LYP based model $\vxc$, and (b) difference between the exact and SCAN0 based model $\vxc$.}}
\label{fig:VXC_benzyne_heatmap_diff}
\end{figure}

Finally, we demonstrate the efficacy of our approach in handling both large (by inverse DFT standards) and strongly correlated systems by conducting a comparative study for the ortho-benzyne molecule ($\text{C}_6\text{H}_4$) in its singlet state. Given the high computational cost for inverse DFT for large molecules, we restrict our comparison on benzyne to B3LYP and SCAN0 functionals. Fig.~\ref{fig:VXC_benzyne_heatmap_exact_b3lyp} and ~\ref{fig:VXC_benzyne_heatmap_exact_scan0} present the error (i.e., $\vxc^{\text{exact}}-\vxc^{\text{model}}$) in the B3LYP and SCAN0 based model $\vxc$, in the plane of the molecule, respectively (refer to Supplementary Material for the individual $\vxc$ potentials). As with LiH and $\text{H}_2\text{O}$, the model $\vxc$ potentials are deeper near the C atoms, highlighting a higher electronegativity on the C atoms compared to the exact $\vxc$. Quantitatively, B3LYP and SCAN0 exhibit similar errors (see Table~\ref{tab:comparison}). The exact and the SCAN0 $\vxc$ exhibit an atomic inter-shell structure around the C atom (see yellow rings around C atoms in Fig. 8 and Fig. 10 in the Supplementary Material), otherwise absent in the B3LYP model $\vxc$.     

\begin{table*}
\caption{\small Comparison of the model exchange-correlation potentials ($\vxc$) in terms of the error metrics $e_1$ and $e_2$ defined in Eq.~\ref{eq:vxcErrors} (cf. the Supplementary Material for $e_3$ and $e_4$ error metrics).}
\begin{tabular}{|p{1.7cm} | M{1.2cm} | M{1.2cm} | M{1.2cm} | M{1.2cm} | M{1.2cm} | M{1.2cm} | M{1.2cm} | M{1.2cm} | M{1.2cm} | M{1.2cm} | M{1.2cm}| M{1.2cm}|}
\hline
\multirow{2}{1.7cm}{Model} & \multicolumn{2}{M{2.4cm}|}{$\text{H}_2(eq)$} & \multicolumn{2}{M{2.4cm}|} {$\text{H}_2(0.8eq)$} & \multicolumn{2}{M{2.4cm}|}{$\text{H}_2(2eq)$} & \multicolumn{2}{M{2.4cm}|}{$\text{LiH}$} & \multicolumn{2}{M{2.4cm}|}{$\text{H}_2\text{O}$}  &
\multicolumn{2}{M{2.4cm}|}{$\text{C}_6\text{H}_4$} \\
\cline{2-13}
&$e_1$&$e_2$& $e_1$ & $e_2$ &  $e_1$ & $e_2$ &  $e_1$ & $e_2$ &  $e_1$& $e_2$ & $e_1$ & $e_2$\\
\hline 
\hline 
B3LYP & 0.192 &  1.657 & 0.216 & 1.125 & 0.272 & 1.377 & 0.107 & 1.862 & 0.041 & 0.309 & 0.049 & 0.071\\ \hline

HSE06 & 0.162 & 0.285 & 0.202 & 0.309 & 0.277 & 0.448 & 0.094 & 1.398 & 0.042 & 0.311 & $-$ & $-$\\ \hline

SCAN0 & 0.145 & 0.240 & 0.181 & 0.167 & 0.273 & 0.302 & 0.087 & 1.010 & 0.030 & 0.227 & 0.045 & 0.098\\ \hline

SCAN & 0.233 & 0.427 & 0.207 & 0.429 & 0.254 & 0.517 & 0.092 & 1.391 & 0.044 & 0.298 & $-$ & $-$\\ \hline

M08-HX & 0.248 & 4.784 &  0.249 & 5.873 & 0.291 & 5.109 & 0.094 & 1.800 & 0.065 & 0.545 & $-$ & $-$\\ \hline

PBE & 0.261 & 1.100 &  0.241 & 0.857 & 0.262 & 1.251 & 0.119 & 2.013 & 0.058 & 0.427 & 0.059 & 0.105\\ \hline

PW92 & 0.288 & 0.267 &  0.267 & 0.278 & 0.297 & 0.420 & 0.152 & 0.264 & 0.133 & 0.354 & 0.145 & 0.407\\ \hline
\end{tabular}
\label{tab:L2DiffWeight}
\end{table*}

In summary, the $\vxc$ of widely used XC functionals differ significantly from those of the exact $\vxc$, for weakly and strongly correlated molecules. Model $\vxc$ potentials exhibit substantial qualitative and quantitative errors, with $\mathcal{O}(10^{-1}-10^{0})$ relative $L_2$ norm errors in XC potentials and gradients. The qualitative differences increase for the stretched $\text{H}_2(2eq)$ molecule in comparison to its equilibrium counterpart, $\text{H}_2(eq)$, highlighting the weakness of current DFT functionals in treating strong correlation. This aspect in particular deserves more attention, and the availability of the exact $\vxc$ may provide important insights. Despite the challenge of strong correlation, SCAN0 and SCAN $\vxc$ give the best overall qualitative agreement with the exact potentials, including the presence of atomic inter-shell structure. Quantitatively, SCAN0 offers the best model among those considered, thereby underscoring the importance of satisfying the known exact conditions as well as incorporating non-local effects of $\vxc$. The ability to compute exact XC potentials via an accurate solution of the inverse DFT problem provides a powerful tool to assess the existing models, as well as enables the possibility of using the exact XC potentials to develop new XC functionals with better accuracy for weakly and strongly correlated systems.

\section*{Acknowledgements}
B.K and V.G. gratefully acknowledge the support of Toyota Research Institute under the auspices of which this line of study was initiated. We acknowledge the support of the Department of Energy, Office of Basic Energy Sciences, grant number DE-SC0022241 under the auspices of which later parts of this work were performed. We also acknowledge the support of the Department of Energy, Office of Basic Energy Sciences, under grant number DE-SC0017320, which supported the computational framework for all-electron calculations essential to this study. This research used resources of the National Energy Research Scientific Computing Center, supported by the Office of Science of the U.S. Department of Energy under Contract No. DE-AC02- 05CH11231. V.G. also acknowledges the support of the Army Research Office through the DURIP grant W911NF1810242, which also provided the computational resources for this work.

%


\onecolumngrid 
\pagebreak
\pagebreak
\begin{center}
\textbf{\Large Supplementary Material}
\end{center}
\section{Errors in model densities and potentials}
We report the error in the densities obtained from self-consistently solved calculations with approximate XC functionals (denoted as $\rhod^{\text{model}}$), relative to the ground-state density from incremental full-CI (iFCI) calculations (denoted as $\rhod^{\text{exact}}$). We quantify the errors in the density using two metrics
\begin{equation}  \label{eq:rhoErrors}
f_1 = \frac{\norm{\rhod^{\text{exact}}-\rhod^{\text{model}}}_{L_2}}{\norm{\rhod^{\text{exact}}}_{L_2}}\,,
\quad f_2 = \frac{\norm{\modulus{\nabla(\rhod^{\text{exact}}-\rhod^{\text{model}})}}_{L_2}}{\norm{\modulus{\nabla\rhod^{\text{exact}}}}_{L_2}}\,.
\end{equation}
Table~\ref{tab:L2DiffRho} lists the $f_1$ and $f_2$ values for all the five benchmark systems considered in this study. We also report two additional error metric for the model $\vxc$'s, given by 
\begin{equation} \label{eq:vxcErrors2}
 e_3 = \frac{\norm{\delta\vxc}_{L_2}}{\norm{\vxc^{\text{exact}}}_{L_2}}\,,\quad e_4 = \frac{\norm{\modulus{\nabla\delta\vxc}}_{L_2}}{\norm{\modulus{\nabla\vxc^{\text{exact}}}}_{L_2}}\,,
\end{equation}
where $\delta\vxc=\vxc^{\text{exact}}-\vxc^{\text{model}}$. We note that while $e_1$ and $e_2$ (presented in the main manuscript) are $\rhod-$weighted error metrics, $e_3$ and $e_4$ are their unweighted counterparts, respectively.  Table~\ref{tab:L2DiffVxc} lists the $e_3$ and $e_4$ error metrics for all the benchmark systems used in this work. Comparing Table~\ref{tab:L2DiffRho} with Table~\ref{tab:L2DiffVxc} (and Table 1 from the main manuscript), we emphasize that while the relative errors in the density are of $\mathcal{O}(10^{-3}-10^{-2})$, the relative errors in the XC potentials are two-orders higher (i.e., $\mathcal{O}(10^{-1}-10^0)$).  In other words, the XC potential exhibits greater sensitivity than the density, and hence, can be instrumental in development of future XC functionals.

\begin{table}[htbp]
\caption{\small Comparing the exact and the model density ($\rho$) in terms of $f_1$ and $f_2$ values (cf. Eq.~\ref{eq:rhoErrors}).} 
\begin{tabular}{|p{1.6cm} | M{1.2cm} | M{1.2cm} | M{1.2cm} | M{1.2cm} | M{1.2cm}| M{1.2cm} | M{1.2cm} | M{1.2cm} | M{1.2cm} | M{1.2cm}|M{1.2cm}| M{1.2cm}|}
\hline
\multirow{1}{1.6cm}{Model} & \multicolumn{2}{M{2.4cm}|}{$\text{H}_2(eq)$} & \multicolumn{2}{M{2.4cm}|} {$\text{H}_2(0.8eq)$} & \multicolumn{2}{M{2.4cm}|}{$\text{H}_2(2eq)$} & \multicolumn{2}{M{2.4cm}|}{$\text{LiH}$} & \multicolumn{2}{M{2.4cm}|}{$\text{H}_2\text{O}$} &
\multicolumn{2}{M{2.4cm}|}{$\text{C}_6\text{H}_4$}\\
\cline{2-13}
& $f_1$ & $f_2 $ & $f_1$ & $f_2$ &  $f_1$ & $f_2$ &  $f_1$ & $f_2$ &  $f_1$ & $f_2$ & $f_1$ & $f_2$ \\
\hline 
\hline
B3LYP & 0.011 & 0.025 &  0.010 & 0.022 & 0.039 & 0.045 & 0.006 & 0.009 & 0.003 & 0.002 & 0.004 & 0.004\\ \hline
 HSE06 & 0.004 & 0.006 & 0.004 & 0.006 & 0.054 & 0.069 & 0.005 & 0.006 & 0.002 & 0.001 & $-$ & $-$\\ \hline
SCAN0 & 0.004 & 0.006 & 0.003 & 0.006 & 0.061 & 0.082 & 0.003 & 0.006 & 0.001 & 0.001 & 0.002 & 0.001\\ \hline
SCAN & 0.006 & 0.012 & 0.005 & 0.011 & 0.043 & 0.054 &  0.004 & 0.008 & 0.002 & 0.001 & $-$ & $-$\\ \hline
M08-HX & 0.018 & 0.047 & 0.019 & 0.049 & 0.062 & 0.080 & 0.006 & 0.011 & 0.002 & 0.003 & $-$ & $-$\\ \hline
PBE & 0.010 & 0.020 & 0.010 & 0.017 & 0.031 &  0.034 & 0.007 & 0.010 & 0.003 & 0.002 & 0.004 & 0.004\\ \hline
PW92 & 0.023 & 0.025 & 0.025 & 0.027 & 0.057 & 0.077 & 0.022 & 0.024 & 0.010 & 0.014 & 0.014 & 0.018\\ \hline
    \end{tabular}
    \label{tab:L2DiffRho}
\end{table}

\begin{table*}
\caption{\small Comparison of the model exchange-correlation potentials ($\vxc$) in terms of the error metrics $e_3$ and $e_4$ (cf.~ Eq.~\ref{eq:vxcErrors2}).}
\begin{tabular}{|p{1.6cm} | M{1.2cm} | M{1.2cm} | M{1.2cm} | M{1.2cm} | M{1.2cm} | M{1.2cm} | M{1.2cm} | M{1.2cm} | M{1.2cm} | M{1.2cm} |M{1.2cm} | M{1.2cm} |}
\hline
\multirow{2}{1.6cm}{Model} & \multicolumn{2}{M{2.4cm}|}{$\text{H}_2(eq)$} & \multicolumn{2}{M{2.4cm}|} {$\text{H}_2(0.8eq)$} & \multicolumn{2}{M{2.4cm}|}{$\text{H}_2(2eq)$} & \multicolumn{2}{M{2.4cm}|}{$\text{LiH}$} & \multicolumn{2}{M{2.4cm}|}{$\text{H}_2\text{O}$} &
\multicolumn{2}{M{2.4cm}|}{$\text{C}_6\text{H}_4$} \\
\cline{2-13}
& $e_3$ & $e_4$ & $e_3$ & $e_4$ & $e_3$ & $e_4$ & $e_3$ & $e_4$ & $e_3$ & $e_4$ & $e_3$ & $e_4$\\
\hline 
\hline 
B3LYP & 0.576 &  0.413 & 0.769 & 0.457 & 0.763 & 0.474 & 0.765 & 0.519 & 0.551 & 0.288 & 0.672 & 0.260 \\ \hline

HSE06 & 0.797 & 1.203 & 0.721 & 0.399 & 0.715 & 0.414 & 0.713 & 0.430 & 0.460 & 0.222 & $-$ & $-$ \\ \hline

SCAN0 & 0.481 & 0.276 & 0.721 &  0.370 & 0.715 & 0.423 & 0.714 & 0.341 & 0.463 & 0.152 & 0.627 & 0.181\\ \hline

SCAN & 0.930 & 0.431 & 0.937 & 0.422 & 0.913 & 0.416 & 0.881 & 0.398 & 0.914 & 0.219 & $-$ & $-$ \\ \hline

M08-HX & 0.451 & 1.319 & 0.460 & 1.383 & 0.460 & 1.448 & 0.443 & 0.886 & 0.436 & 0.684 & $-$ & $-$\\ \hline

PBE & 0.927 & 0.472 & 0.935 & 0.487 & 0.912 & 0.485 & 0.873 & 0.536 & 0.907 & 0.311 & 0.274 & 1.213  \\ \hline

PW92 & 0.927 & 0.437 & 0.934 & 0.438 & 0.912 & 0.435 & 0.875 & 0.484 & 0.910 & 0.334 & 0.821 & 0.340 \\ \hline
\end{tabular}
\label{tab:L2DiffVxc}
\end{table*}

\section{Comparison of XC potentials}
\begin{figure}
    \centering
    \includegraphics[scale=0.7]{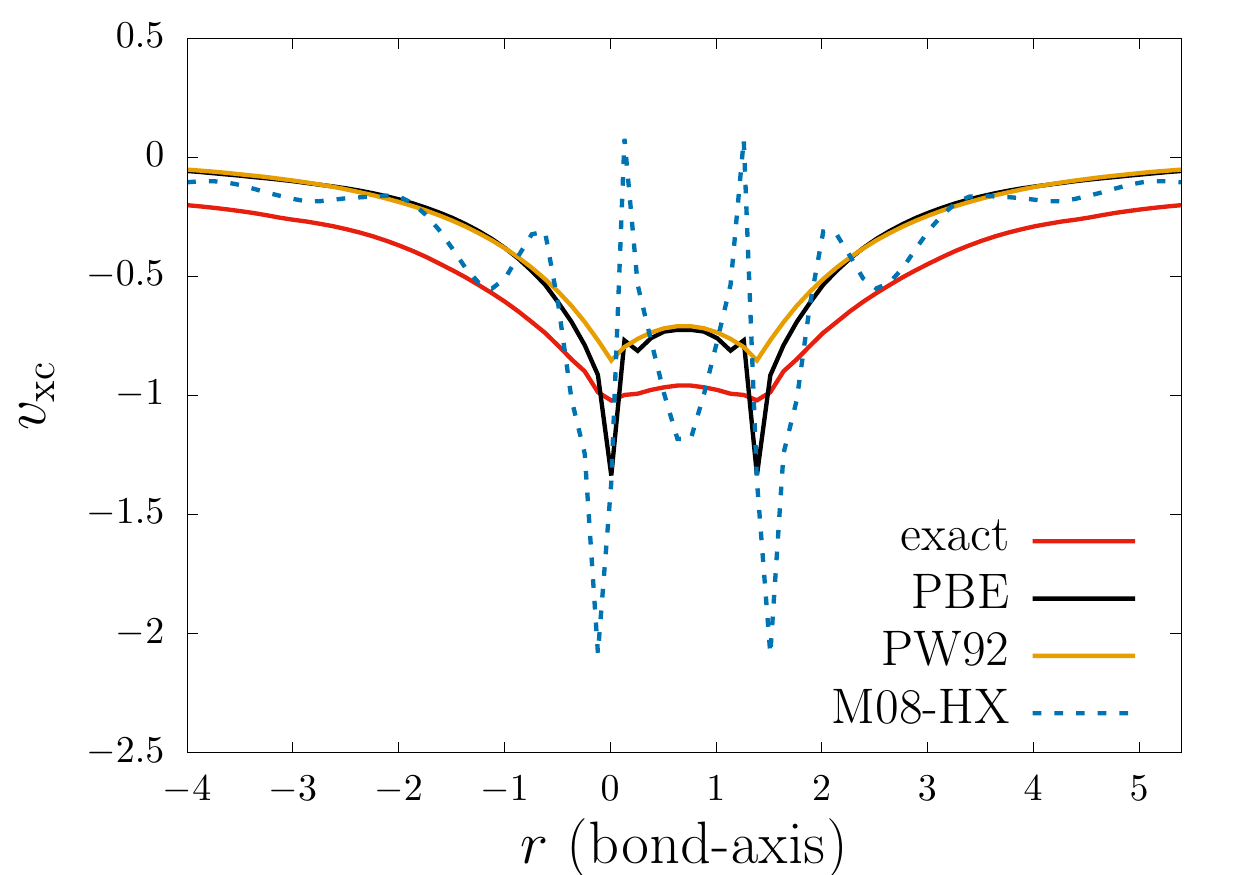}
    \caption{\small Comparison of the exact, M08-HX, PBE, and PW92 based $\vxc$ for $\text{H}_2(eq)$.}
     \label{fig:VXC_H2eq_SI}
\end{figure}

\begin{figure} 
    \centering
    \includegraphics[scale=0.7]{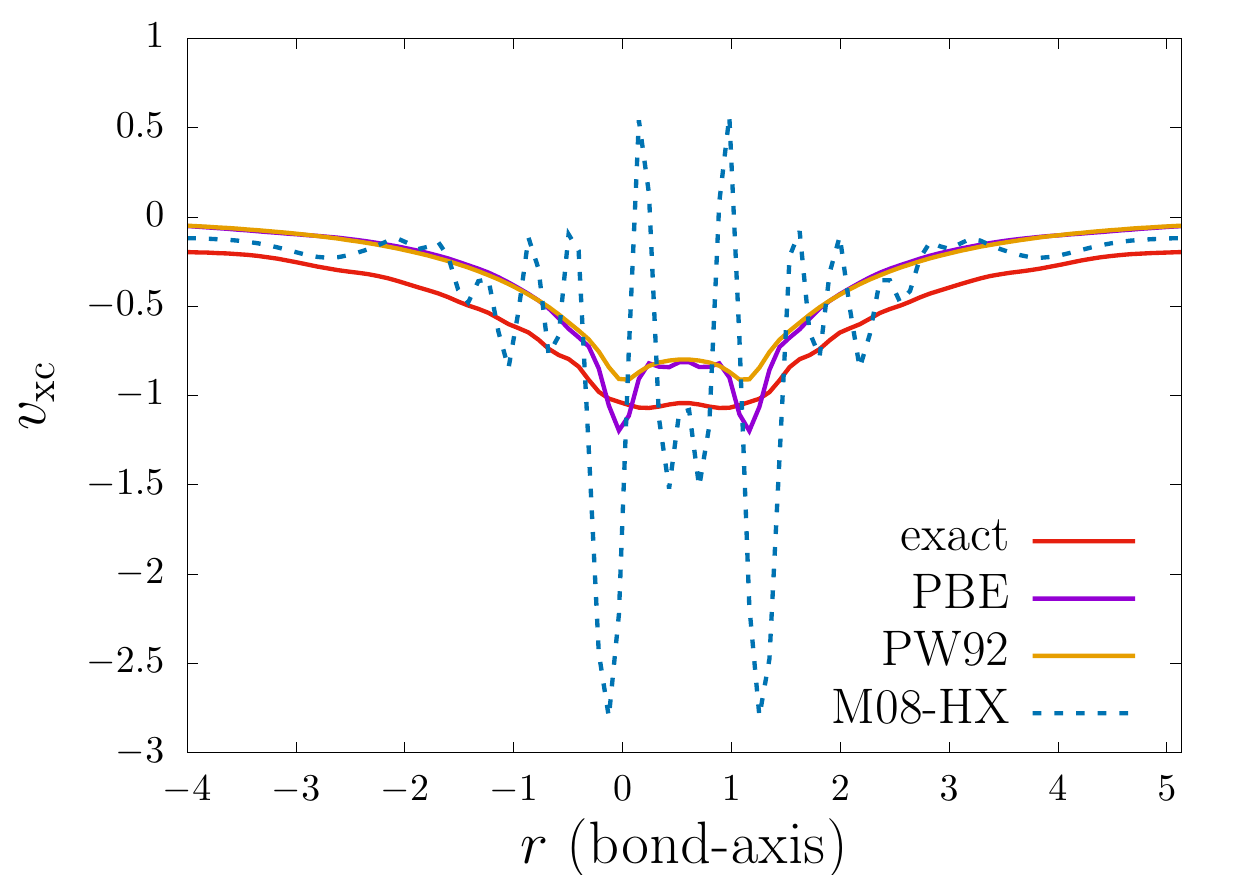}
    \caption{\small Comparison of the exact, M08-HX, PBE, and PW92 based $\vxc$ for $\text{H}_2(0.8eq)$.}
    \label{fig:VXC_H20.8eq_SI}
\end{figure}

\begin{figure} 
    \centering
    \includegraphics[scale=0.7]{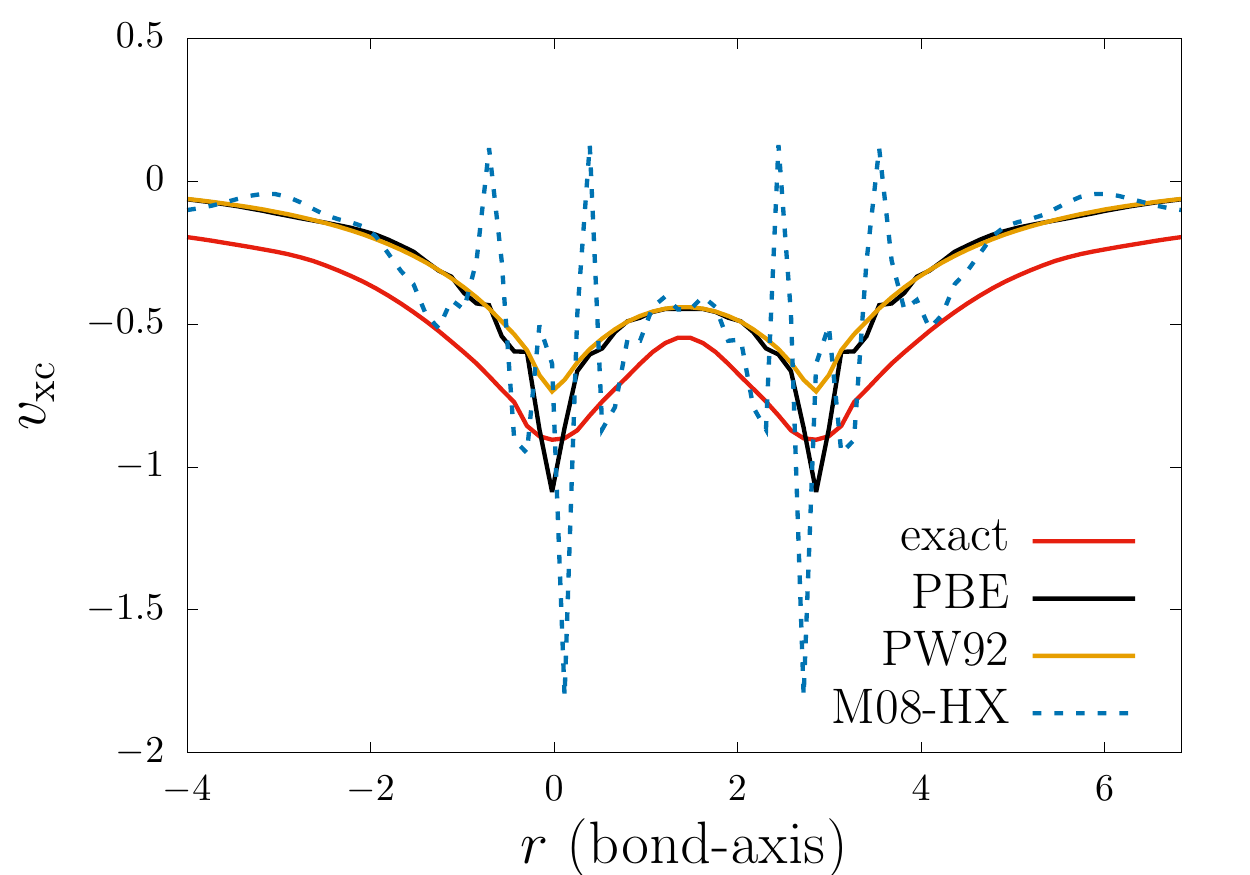}
    \caption{\small Comparison of the exact, M08-HX, PBE, and PW92 based $\vxc$ for $\text{H}_2(2eq)$.}
    \label{fig:VXC_H22eq_SI}
\end{figure}

\begin{figure}
    \centering
    \includegraphics[scale=0.7]{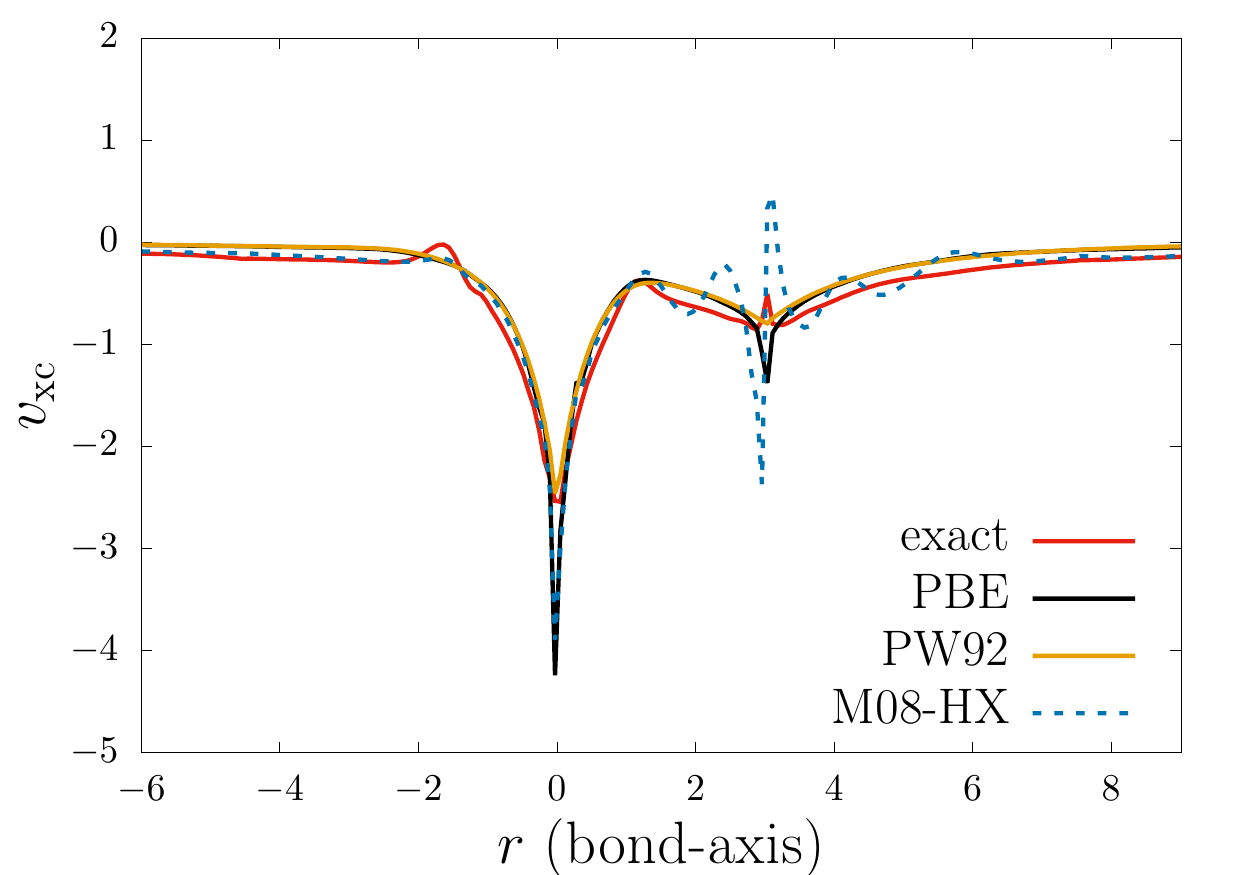}
    \caption{\small Comparison of the exact, M08-HX, PBE, and PW92 based $\vxc$ for $\text{LiH}$.}
    \label{fig:VXC_LiH_SI}
\end{figure}

\begin{figure}
    \centering
    \includegraphics[scale=0.7]{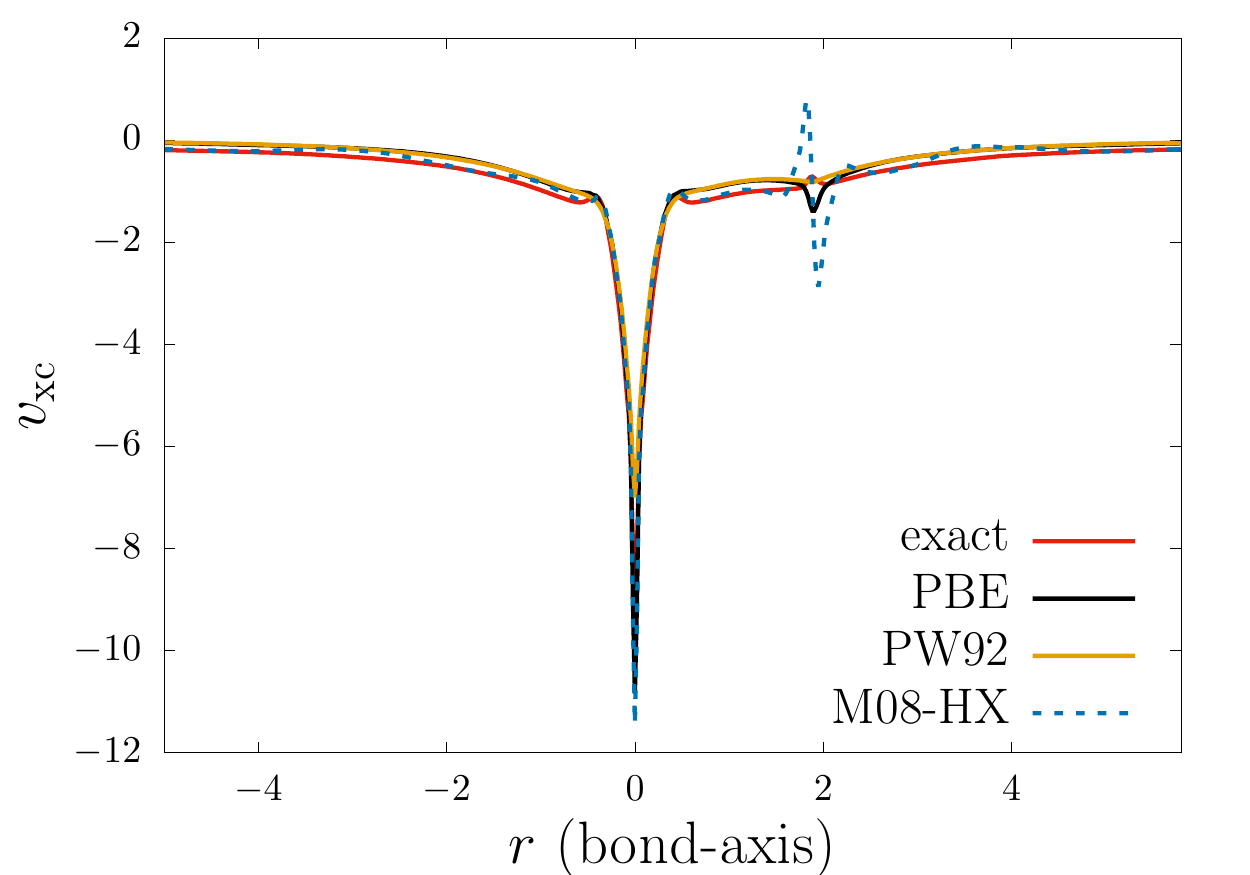}
    \caption{\small Comparison of the exact, M08-HX, PBE, and PW92 based $\vxc$ for $\text{H}_2\text{O}$.}
    \label{fig:VXC_H2O_bond_SI}
\end{figure}

In this section, we present a comparison of the exact $\vxc$ against M08-HX, PBE, and PW92 based model $\vxc$ for all the five benchmark systems considered in this work. Figs.~\ref{fig:VXC_H2eq_SI},~\ref{fig:VXC_H20.8eq_SI},~\ref{fig:VXC_H22eq_SI}, ~\ref{fig:VXC_LiH_SI}, and ~\ref{fig:VXC_H2O_bond_SI}  present the comparison for $\text{H}_2(eq)$, $\text{H}_2(0.8eq)$, $\text{H}_2(2eq)$, $\text{LiH}$, and $\text{H}_2\text{O}$, respectively. As evident, the model $\vxc$ differ significantly from the exact $\vxc$. The PBE and PW92 based $\vxc$, on account of being local functionals, rapidly decay to zero in the far-field, as opposed to the $-1/r$ decay in the exact one. The PBE based $\vxc$ is substantially deeper at the atoms, compared to the exact $\vxc$, owing to a greater concentration of density around the atoms. The M08-HX based $\vxc$ exhibit large unphysical oscillations, possibly, due to serious errors in treatment of the exchange potential~\cite{Mardirossian2013,Medvedev2017}. For the LiH and $\text{H}_2\text{O}$, the M08-HX, PBE, and PW92 lack the atomic intershell structure around the Li and the O atom, respectively, otherwise present in the exact $\vxc$. Further, both the PBE and PW92 based $\vxc$ exhibit a local minima at the H atom in LiH and $\text{H}_2\text{O}$, as opposed to a local maxima in the exact case.  

\begin{figure}
    \centering
    \includegraphics[scale=0.7]{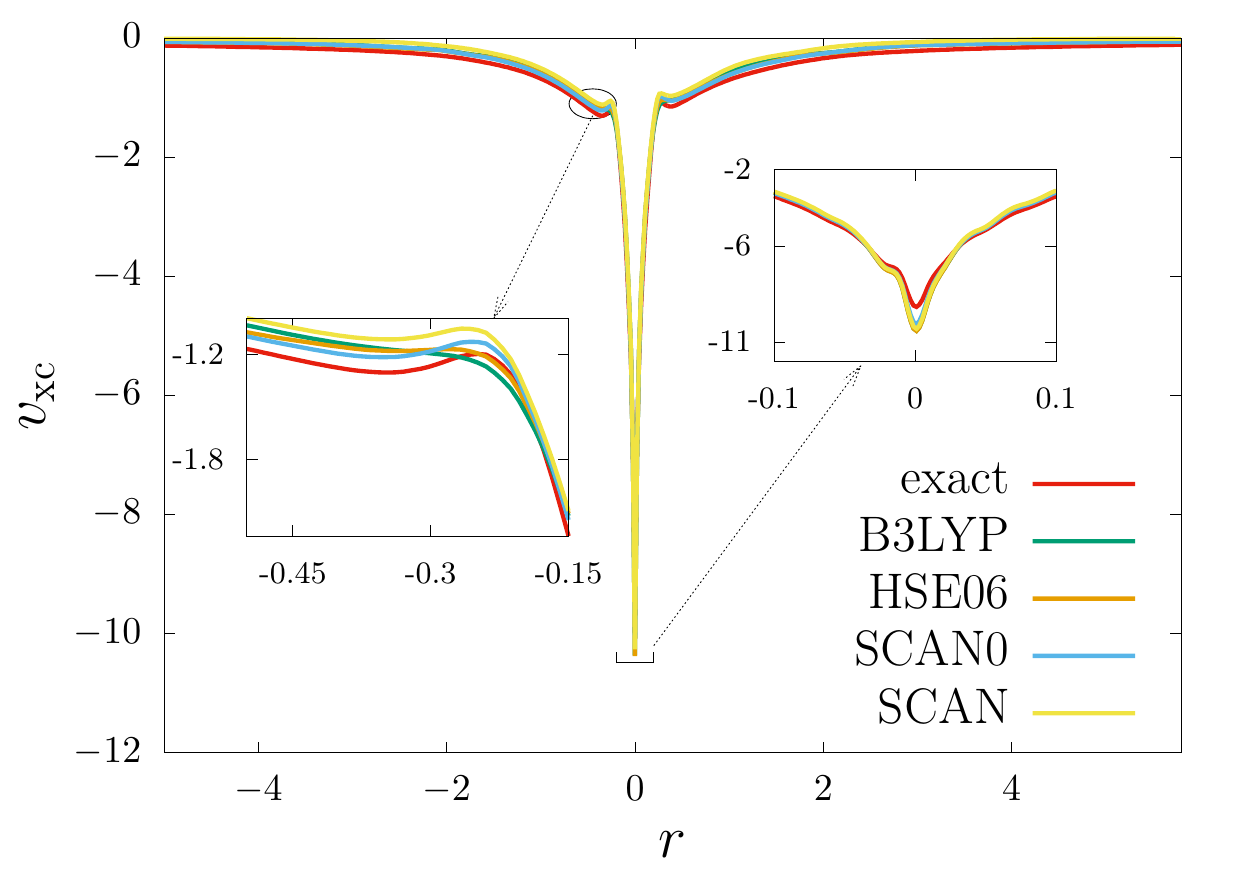}
    \caption{\small Comparison of the exact, B3LYP, HSE06, SCAN0 and SCAN based $\vxc$ for $\text{H}_2\text{O}$ along the lone-pair axis (i.e., bisector of the H-O-H angle). The O atom is at $r=0$.}
    \label{fig:VXC_H2O_lone_pair}
\end{figure}

\begin{figure}[htbp]
\centering
\subfigure[]{ \label{fig:VXC_H2O_bond_error}\includegraphics[scale=0.7]{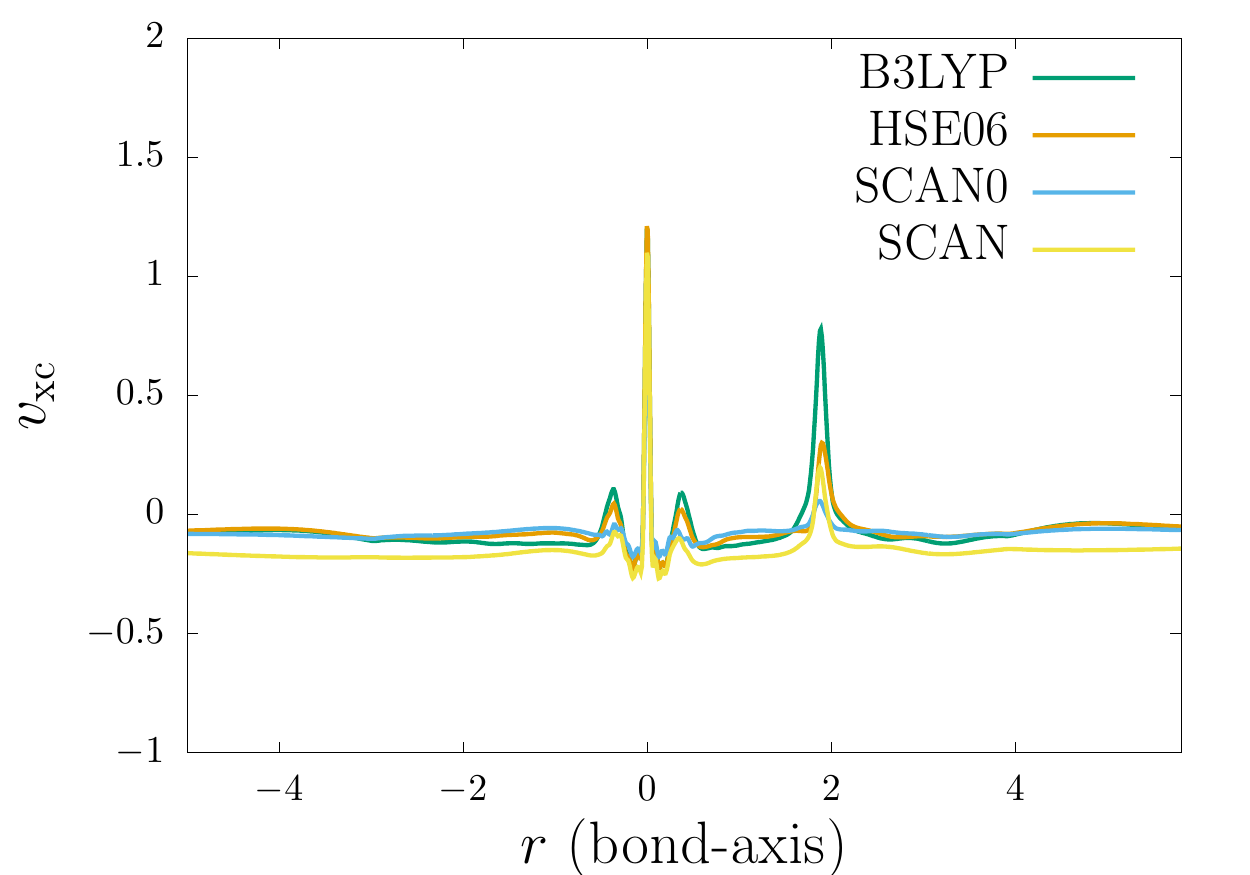}}
\vspace{-0.3cm}
\subfigure[]{ \label{fig:VXC_H2O_lone_pair_axis_error}\includegraphics[scale=0.7]{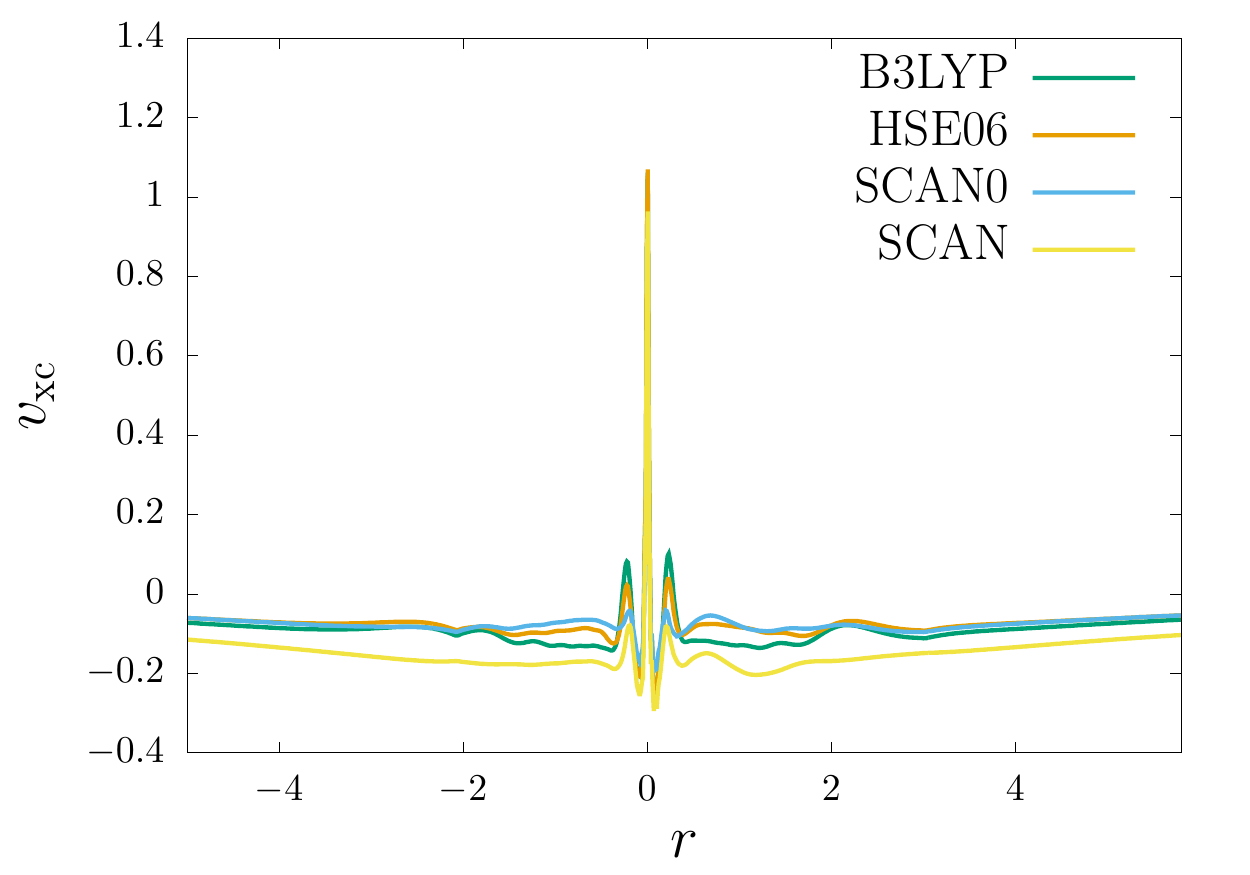}}
\caption{\small{Difference between the exact and model $\vxc$ (i.e., $\vxc^{\text{exact}}-\vxc^{\text{model}}$) for the $\text{H}_2{\text{O}}$ molecule along: (a) the O-H bond, and (b) the lone-pair axis.}}
\label{fig:VXC_H2O_error}
\end{figure}

\begin{figure}[htbp]
\centering
\includegraphics[scale=0.2]{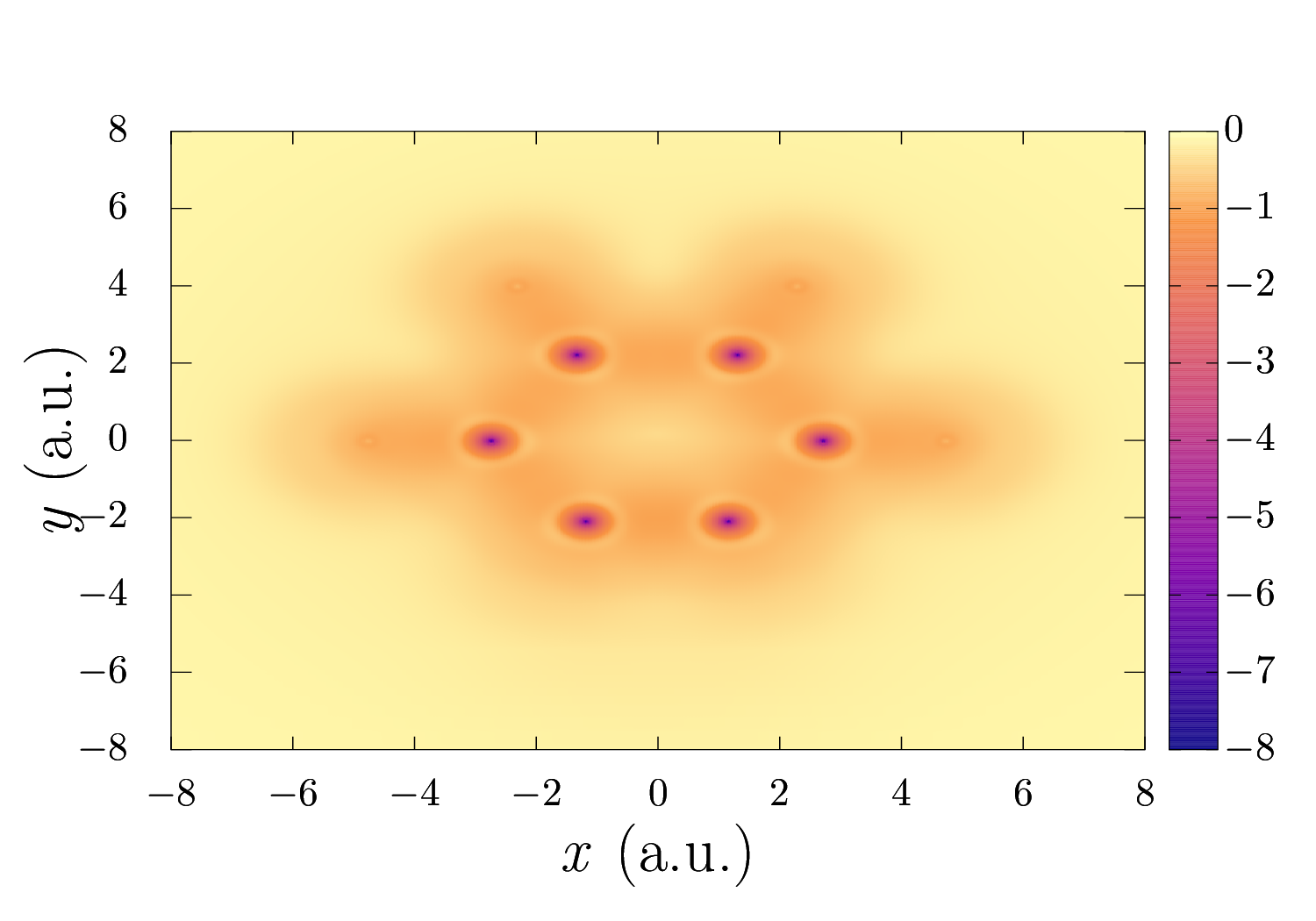}
\caption{\small{Exact $\vxc$ for benzyne ($\text{C}_6\text{H}_4$) in the plane of the molecule.}}
\label{fig:VXC_benzyne_exact}
\end{figure}

\begin{figure}[htbp]
\centering
\includegraphics[scale=0.2]{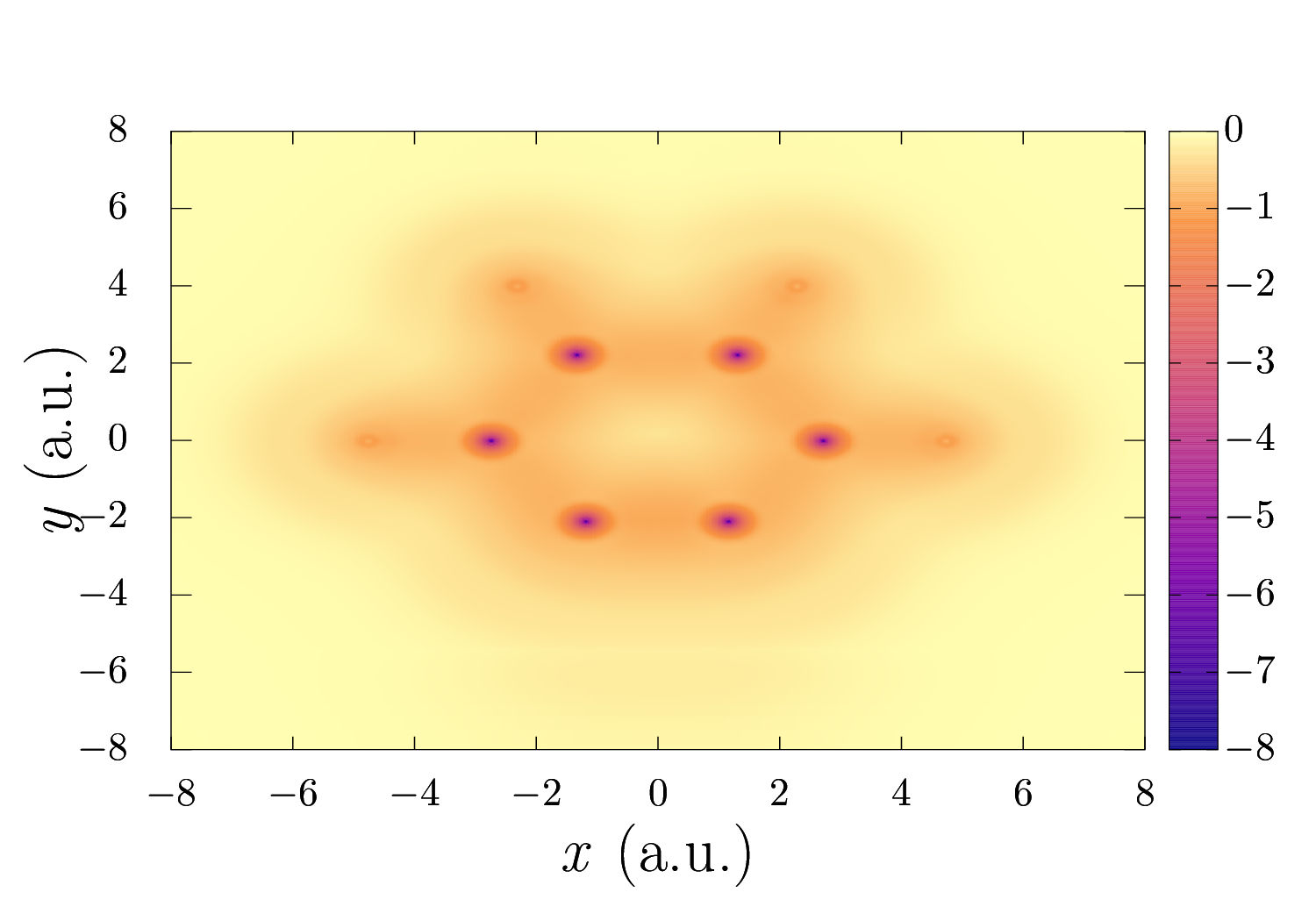}
\caption{\small{B3LYP based model $\vxc$ for benzyne ($\text{C}_6\text{H}_4$) in the plane of the molecule.}}
\label{fig:VXC_benzyne_b3lyp}
\end{figure}

\begin{figure}[htbp]
\centering
\includegraphics[scale=0.2]{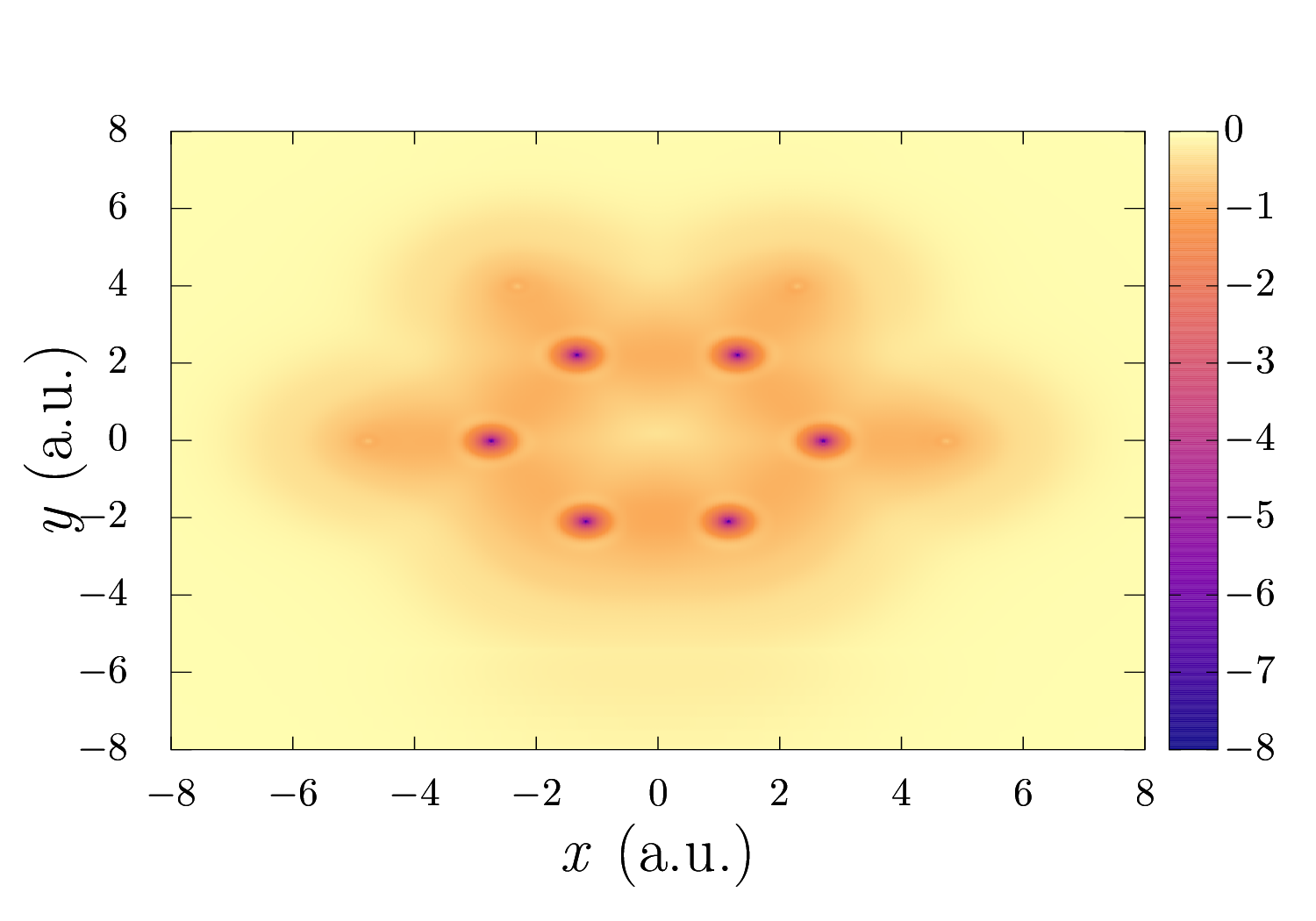}
\caption{\small{SCAN0 based model $\vxc$ for benzyne ($\text{C}_6\text{H}_4$) in the plane of the molecule.}}
\label{fig:VXC_benzyne_scan0}
\end{figure}

In Fig.~\ref{fig:VXC_H2O_lone_pair}, we present a comparison of the exact $\vxc$ and the model $\vxc$ (B3LYP, HSE06, SCAN0, and SCAN based) for $\text{H}_2\text{O}$ along the lone-pair axis (i.e., along the angular bisector of the H-O-H angle). Fig.~\ref{fig:VXC_H2O_error} provides the error in the model $\vxc$ along both the O-H bond as well as the lone-pair axis. Lastly, Fig.~\ref{fig:VXC_benzyne_exact}, Fig.~\ref{fig:VXC_benzyne_b3lyp}, and Fig.~\ref{fig:VXC_benzyne_scan0} provides the exact, the B3LYP-based, and the SCAN0-based $\vxc$ for ortho-benzyne ($\text{C}_6\text{H}_4$).

\end{document}